# A Study of Dynamic Stock Relationship Modeling and S&P500 Price Forecasting Based on Differential Graph Transformer

Linyue Hu, Qi Wang*

*China Agricultural University, Beijing, China*

**Abstract**:

Stock price prediction is vital for investment decisions and risk management, yet remains challenging due to markets' nonlinear dynamics and time-varying inter-stock correlations. Traditional static-correlation models fail to capture evolving stock relationships. To address this, we propose a Differential Graph Transformer (DGT) framework for dynamic relationship modeling and price prediction.

Our DGT integrates sequential graph structure changes into multi-head self-attention via a differential graph mechanism, adaptively preserving high-value connections while suppressing noise. Causal temporal attention captures global/local dependencies in price sequences. We further evaluate correlation metrics (Pearson, Mutual Information, Spearman, Kendall's Tau) across global/local/dual scopes as spatial-attention priors.

Using 10 years of S&P 500 closing prices (z-score normalized; 64-day sliding windows), DGT with spatial priors outperformed GRU baselines (RMSE: 0.24 vs. 0.87). Kendall's Tau global matrices yielded optimal results (MAE: 0.11). K-means clustering revealed "high-volatility growth" and "defensive blue-chip" stocks, with the latter showing lower errors (RMSE: 0.13) due to stable correlations. Kendall's Tau and Mutual Information excelled in volatile sectors.

This study innovatively combines differential graph structures with Transformers, validating dynamic relationship modeling and identifying optimal correlation metrics/scopes. Clustering analysis supports tailored quantitative strategies. Our framework advances financial time-series prediction through dynamic modeling and cross-asset interaction analysis.

**Keywords**：Differential Graph Transformer; Dynamic Stock Relationship Modeling; Stock Price Prediction; Interpretability Analysis

* **Corresponding author**: College of Science, China Agricultural University, Beijing, 100083, China.
Email Address: wangqi_math@cau.edu.cn (Q. Wang)

# 1. Introduction

## 1.1 Research background and significance

As more individuals and institutions around the world turn to the stock market for asset appreciation, it becomes increasingly apparent that actively managed equity funds struggle to outperform their benchmarks. Taking the U.S. market as an example, in a large-cap-dominated investment environment, most actively managed stock funds still fail to beat their benchmarks. As shown in Figure 1, according to the latest SPIVA U.S. Scorecard report published by S&P Dow Jones Indices (Ganti et al., 2024), as many as 65% of actively managed large-cap equity funds underperformed the S&P 500 Index in 2024, meaning that only about 35% of funds outperformed the benchmark.

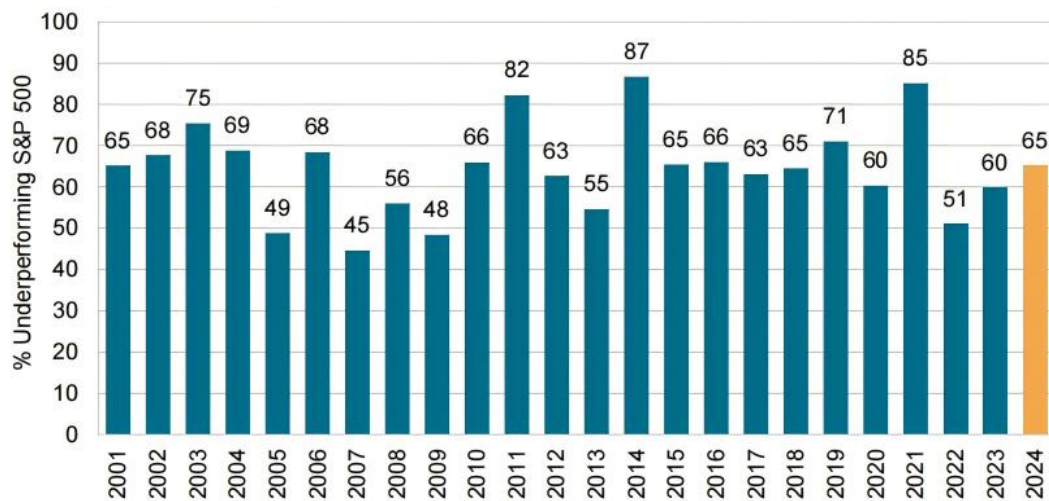

**Figure 1 Annual percentage of U.S. large-cap funds underperforming the S&P 500 from 2001 to 2024**
**Source(s):** S&P Dow Jones Indices LLC, CRSP. Data as of Dec.31, 2024. Past performance is no guarantee of future results. Chart is provided for illustrative purposes.

More notably, the difficulty of outperforming the market does not significantly improve over longer time horizons. Over the 15-year period ending in December 2024, the majority of active funds across all stock segments (including large-, mid-, and small-cap stocks) also failed to consistently beat their respective benchmarks (Ganti et al., 2024). This persistent underperformance highlights the formidable challenge of finding stable alpha-generating models and strategies in a stock market characterized by noise and evolving interrelationships.

Similarly, the performance of actively managed funds in China's A-share market has been suboptimal. With regulatory authorities advocating for index-based investing, domestic equity index

ETFs experienced record net inflows in 2024, reflecting investors' growing preference for low-cost, passive products. In contrast, actively managed funds continue to lose assets due to their persistent underperformance. According to a Reuters report, the CSI Active Equity Fund Index (.CSI930890) rose by only 3% in 2024, significantly lagging behind the 16% gain of the benchmark CSI 300 Index ("China Funds Slash ETF Fees, Escalating Price War in Booming Market", 2024). These data clearly show that, whether in the mature U.S. market or the rapidly developing Chinese market, the vast majority of active investors struggle to consistently outperform the market.

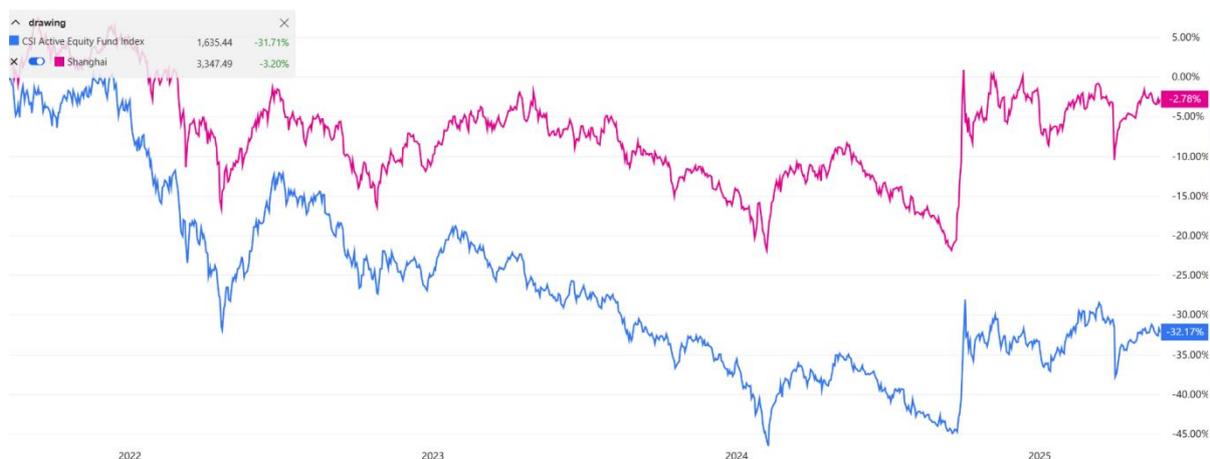

**Figure 2 Comparison of return volatility between the CSI Active Equity Fund Index and the S&P 500 over the past three years**

However, "beating the index" is no easy task. Stock prices are influenced by a complex mix of macroeconomic conditions, industry cycles, company fundamentals, and market sentiment, displaying high levels of non-linearity and noise. At the same time, the market structure evolves over time, with emerging industries rising and the weighting of large tech companies continuing to increase—posing significant challenges for traditional linear or low-dimensional models. Additionally, short-term volatility is often dominated by random events, making classical statistical or machine learning approaches prone to overfitting or underfitting, thereby limiting their predictive accuracy in real-world applications.

If a model can be developed that captures both the dynamic relationships among stocks and the temporal evolution of their prices, it would offer significant practical value to investors. First, more accurate price forecasts can contribute to better portfolio construction and reduced tail risks. Second, modeling the dynamic inter-stock relationships helps identify sector rotation and structural opportunities in a timely manner, supporting asset allocation and timing strategies. Lastly, from a macro perspective, such models can assist regulators in identifying systemic risk transmission pathways, promoting the stability and healthy development of financial markets.

Based on this motivation, this study attempts to introduce a Differential Graph Structure and leverage the powerful temporal modeling capabilities of the Transformer, with the aim of making breakthroughs in modeling the dynamic relationships among S&P 500 constituent stocks—ultimately providing valuable insights for active investors seeking to outperform the market.

## 1.2 Research Objectives

This study aims to develop a dynamic stock relationship modeling framework based on the Differential Graph Transformer (DGT) to enhance the predictive power for S&P 500 constituent stock prices and to systematically evaluate how different methods of modeling inter-stock correlation affect predictive performance. The specific objectives include:

**(1) Capturing dynamic relationships among stocks**

Construct and dynamically update a differential graph structure to reflect the non-stationary correlations among constituent stocks as they evolve over time. The differential graph construction mechanism is designed to more accurately capture structural changes driven by sector rotation, market style shifts, and unexpected events.

**(2) Integrating Transformer and graph structures for temporal modeling**

Embed historical price sequence information within the graph structure and utilize the Transformer's multi-head attention mechanism to simultaneously capture local temporal trends and cross-node influences among stocks. This approach aims to enhance the model's sensitivity to price fluctuations and improve its overall stability.

**(3) Comparing model performance under different correlation matrices**

Build input graphs based on various correlation measures—such as Pearson, Mutual Information, Spearman, and Kendall's Tau—and construct global, local, and hybrid graph structures. Evaluate how each graph construction method affects the DGT model's predictive performance, and explore the applicability and advantages of different correlation metrics in dynamic relationship modeling.

**(4) Evaluating predictive performance and practical value**

Backtest the model using historical market data and assess its predictive accuracy and robustness using RMSE and MAE metrics. Additionally, through case studies, analyze the model's performance across different industries and volatility regimes to evaluate its practical value and potential limitations in portfolio construction.

By achieving these goals, this study aims to provide an interpretable and adaptive solution for modeling dynamic relationships and forecasting high-dimensional time series in the stock market, thereby serving as a forward-looking support tool for quantitative investment strategies.

## 1.3 Literature Review

Financial market forecasting has long been a key interdisciplinary area of research between finance and computer science. With the development of deep learning technologies—especially the rise of Graph Neural Networks (GNNs) and Transformer models—new perspectives have emerged for stock market prediction. Early stock forecasting methods primarily relied on time series analysis techniques, such as ARIMA and GARCH statistical models (Zhang et al., 2017). While these methods were somewhat effective in capturing the temporal characteristics of individual stocks, they struggled to model the complex interrelationships among different stocks.

GNNs offer a new tool for modeling such inter-stock relationships. Chen et al. (2018) employed Graph Convolutional Networks (GCNs) to integrate inter-company relationships into stock price prediction models, using node embeddings to learn distributed representations of each firm. By capturing the network structure and mutual influence among companies, they improved prediction accuracy. Cheng et al. (2022) developed a Multi-modal Attention-based Graph Neural Network (MAGNN) that applies a two-stage attention mechanism for joint optimization, learning from multimodal inputs to perform financial time series prediction. Zheng et al. (2023) proposed a Relation-Time Graph Convolutional Network (RT-GCN), which constructs a multi-relational temporal graph and fuses graph convolutions with time series models to achieve ranking-based stock prediction. Their method showed significant performance gains on the S&P 500 and NASDAQ datasets.

Transformer models (Vaswani et al., 2017), known for their powerful parallel computing and long-range dependency modeling capabilities, have opened new opportunities for analyzing financial time series data. Ding et al. (2020) proposed an enhanced Transformer for stock trend forecasting, incorporating innovations such as multi-scale Gaussian priors, orthogonal regularization, and trading-interval segmentation to improve the standard Transformer's ability to capture long-term dependencies in financial time series. Wang et al. (2022) applied Transformers to forecast stock market indices and conducted multiple back-testing experiments on major global indices, demonstrating that Transformers significantly outperformed other methods. Li et al. (2023) proposed the FDGRU-Transformer model, which uses empirical mode decomposition to split noisy stock data into trend and independent components, and combines GRU, LSTM, and multi-head attention to preserve fine-grained temporal information—enabling the model to extract more discriminative features from chaotic market data.

Integrating GNNs and Transformers is becoming a promising research direction. Xiang et al. (2022) proposed the THGNN model to learn dynamic relations of price movements in financial time series. By constructing company relation graphs based on historical price dynamics and using a Transformer encoder for temporal representation, they demonstrated superior performance across U.S. and Chinese stock markets. Qian et al. (2024) introduced a Multi-relational Dynamic Graph Neural Network (MDGNN) framework that captures the temporal evolution of stock interrelations via discrete dynamic graphs, and integrates Transformer-based temporal encoding to predict stock investments—achieving state-of-the-art performance on public datasets. Moghimi et al. (2023) extended the Transformer-GNN (T-GNN) approach to the real estate pricing domain and proposed an end-to-end, data- and preprocessing-independent solution that improved performance by nearly 50% over traditional methods, while also offering interpretable property valuation insights.

## 1.4 Main Contributions and Innovations

This paper aims to enhance the modeling of the dynamic structure of the stock market to better support constituent stock price forecasting. In contrast to existing GNN-based approaches that heavily rely on static or manually sparsified correlation graphs, we propose a Transformer model based on a Differential Graph Attention Mechanism (DGT), which dynamically constructs and updates graph structures to better capture time-varying correlations and structural shifts in financial markets.

Unlike traditional methods that directly build correlation graphs, DGT introduces a differential graph attention mechanism. It models the difference between "base attention weights" and "enhanced attention weights" to adaptively retain the most informative edges for prediction, while suppressing redundant or noisy connections. This generates sparse and stable dynamic graph structures that adjust automatically over time — without requiring additional threshold hyperparameters or fixed top-k strategies—thus offering greater flexibility and interpretability.

Moreover, we systematically incorporate multiple correlation metrics (Pearson, Mutual Information, Spearman, Kendall's Tau) to construct global, local, and purely dynamic graphs. These are ultimately integrated via multi-head attention, balancing long-term trends with short-term fluctuations and enhancing the expressiveness of graph structures across different temporal scales. On top of this, the model fuses causal temporal attention with differential graph attention to jointly model price sequences and their time-varying graph relationships—effectively capturing complex temporal and inter-stock dependencies in financial time series.

For model implementation, we construct a forecasting task based on ten years (from 2015 to 2025) of historical S&P 500 constituent stock data. A rolling window mechanism is employed for thorough backtesting, and model performance is evaluated using RMSE and MAE metrics. Additionally, case studies across different industry sectors and volatility levels demonstrate the model's adaptability under various market conditions. We also explore its potential applications and limitations in practical portfolio construction.

## 1.5 Thesis Structure

This chapter introduces the research background and significance, clarifies the objectives and motivations, reviews key related literature, and outlines the main contributions and innovations of this study.

The Related Works chapter surveys traditional stock price prediction methods, the application of GNNs in finance, and the latest advancements in Transformer models and differential attention mechanisms, laying a theoretical foundation for the proposed method.

The Methodology chapter provides a detailed description of the DGT architecture, including input projection, causal temporal attention, and differential graph attention modules. It also discusses learning rate choices, grid search tuning strategies, loss functions, and evaluation metrics.

The Research process and Results chapter presents the dataset sources and preprocessing steps, experimental settings and parameter configurations, and the model's backtesting results on historical market data, evaluated using RMSE and MAE.

The Discussion chapter further analyzes the results by clustering stocks and comparing the DGT model's performance across different stock groups.

The Conclusion section summarizes the key findings, discusses the study's limitations, and proposes directions for future work in model improvement, data expansion, and real-world applications.

## 2. Related works

### 2.1 Traditional Stock Prediction Methods

In early research on stock price prediction, scholars mainly relied on statistical and classical machine learning models. These methods can generally be divided into two categories: time-series-based models and supervised learning based on feature engineering.

First, time-series models such as ARIMA, GARCH, and their variants were the most commonly used tools. The ARIMA model captures linear dependencies in the sequence through autoregressive and moving average components but requires the data to be stationary and can only fit linear relationships. GARCH and its extensions can model volatility clustering to some extent but struggle to handle nonlinearity or multiple jump behaviors in the series. Moreover, these models are highly sensitive to parameter settings and window lengths, often failing when faced with rapidly changing market structures.

Another major category includes supervised learning methods based on feature engineering. Researchers extract hundreds or even thousands of features from technical indicators (e.g., moving averages, relative strength index), fundamental indicators (e.g., P/E ratio, P/B ratio), and sentiment indicators (e.g., news sentiment scores), then apply algorithms such as linear regression, support vector machines, decision trees, random forests, and XG Boost for training. These methods can learn nonlinear mappings and offer more flexibility than pure statistical models. However, their predictive accuracy heavily depends on the quality and diversity of features. They often require extensive manual tuning and refined feature selection and are prone to overfitting in high-dimensional spaces, making it difficult to generalize to new scenarios.

In recent years, deep learning models such as feedforward neural networks (MLP), convolutional neural networks (CNN), recurrent neural networks (RNN), and long short-term memory networks (LSTM) have been widely applied in stock prediction. Through multi-layer nonlinear transformations and automatic feature learning, they can mine deep patterns in time-series data to some extent. However, these models generally suffer from the "black-box" drawback, lack the ability to capture interactions between different stocks, and are more dependent on hyperparameters and large training datasets. In noisy, small-sample financial time-series contexts, deep models are prone to local optima or even catastrophic forgetting without effective regularization or structural priors.

Overall, traditional stock prediction methods either focus on modeling linear or weakly nonlinear dynamics of single sequences or rely on handcrafted features to characterize individual stocks. However, they generally overlook dynamic inter-stock relationships and information propagation mechanisms. These inherent limitations leave ample room for research based on graph neural networks (GNNs) and Transformers to model cross-asset relationships.

## 2.2 Graph Neural Networks and Stock Prediction

Due to fundamental and sentiment factors, stocks often exhibit common trends. However, different stocks also show unique behaviors due to their specific characteristics. As shown in Figure 3, the price trends of Apple, Microsoft, and Google over the past ten years demonstrate both similarities and distinct patterns.

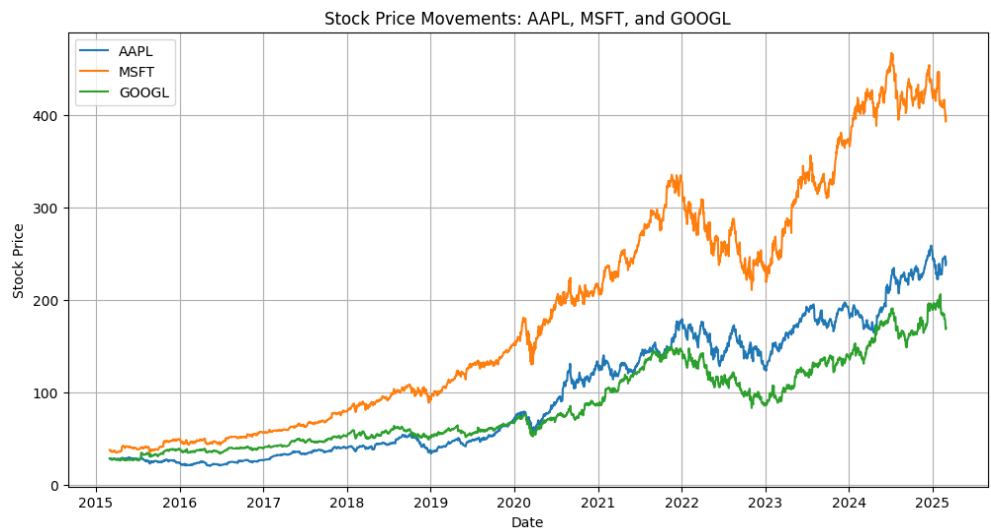

**Figure 3 Ten-Year Stock Price Trends of Apple, Microsoft, and Google**

However, stocks are not isolated entities. Stocks within the same industry or with similar attributes tend to be correlated. This means that in response to favorable news or shocks, their prices may move in the same or opposite directions. Such information is often ignored by traditional prediction methods.

With the success of GNNs in fields like social networks and recommender systems, researchers have begun exploring their application in stock market prediction to capture the complex dependencies among component stocks. When constructing stock graphs, common approaches include company relationship graphs, text-based graphs, and statistical graphs. Among these, statistical correlation graphs are widely used due to their ease of construction and strong timeliness based solely on historical price data. A typical method is to calculate the correlation coefficients between stock pairs to generate a fully connected weighted graph. However, such fully connected graphs are often overly dense, leading to over-smoothing in GNNs. They require sparsification via thresholding or top-k selection, which can discard correlation strength information and is highly sensitive to hyperparameter choices.

To address the issue of static global correlation graphs failing to capture dynamic market changes, Yin et al. (2021) were the first to apply global Pearson correlation graphs on the Dow Jones and ETF datasets, combining GCN and GRU models for price prediction. In this framework, GCN generates node embeddings at each time step, while GRU captures temporal dependencies, significantly outperforming GRU-only baselines. However, the fixed graph structure limits the model's adaptability to sector rotation and style shifts, thus constraining further improvements in prediction performance.

To better accommodate long-term trends and short-term fluctuations, Ma et al. (2024) proposed a method that fuses global and local Pearson correlation graphs in a multi-graph convolutional network. This method was validated on DJIA, ETF, and SSE datasets and proved effective. Although this approach mitigates the limitations of static graphs, it still relies on thresholding for sparsity and predefined correlations, which may not reflect more complex nonlinear interactions in the market.

While the above methods have partially improved the static and dense nature of graph structures, they still rely on manually set hyperparameters for sparsification and multi-graph fusion, making it difficult to adapt to all stocks and market conditions. Against this background, the Differential Graph Transformer was introduced. It aims to dynamically adjust edge weights through a differential attention mechanism while simultaneously integrating multi-scale priors, making graph construction and updating more flexible and efficient.

## 2.3 Transformer and Differential Attention

As previously discussed, in financial time series modeling, stock prices are influenced not only by their own historical trends but also by complex, dynamic interactions with other stocks. Traditional methods based on fixed graph structures or static time series struggle to effectively capture these time-varying cross-asset relationships. To address this, this paper introduces the Transformer framework to extract temporal dependencies, and further proposes a differential graph-based attention mechanism to model the dynamically evolving inter-stock relationships over time.

The Transformer is a deep neural network architecture based on the self-attention mechanism, originally developed for natural language processing tasks. In recent years, it has been widely applied to time series forecasting and financial modeling. Unlike RNNs and other sequential models, the Transformer can model dependencies between any positions in parallel, significantly improving modeling efficiency and flexibility.

Its core components include:

**(1) Scaled Dot-Product Attention**

Given matrices of queries $Q$, keys $K$, and values $V$, the attention weights are computed as:

$$Attention(Q, K, V) = softmax(\frac{QK^T}{\sqrt{d_k}})V$$

where $d_k$ is the dimension of the key vectors. This mechanism measures similarity through inner products and generates weighted combinations of the values.

**(2) Multi-Head Attention**

The model maps $Q$, $K$, and $V$ into multiple subspaces and computes attention in parallel, allowing it to capture multi-level dependency structures.

**(3) Positional Encoding**

To compensate for the lack of sequence awareness in the architecture, positional encoding vectors are added to the input embeddings. These encode both relative and absolute position information in time.

This architecture provides natural advantages for modeling high-dimensional and dynamic financial time series, particularly when uncovering complex dependencies across assets and time scales. However, a limitation of the standard Transformer is its assumption that relationships between input variables are temporally static. In reality, inter-stock dependencies are not fixed — they evolve continuously due to macroeconomic conditions, sector dynamics, market sentiment, and other factors. As a result, the standard Transformer struggles to capture the temporal evolution of relational structures.

To overcome this, we introduce a differential attention mechanism.

Specifically, instead of directly modeling the stock relation graph at a single time point, we focus on the change in graph structure over consecutive time steps, in order to extract dynamic signals that reflect shifts in market structure. If we denote the stock relation graphs at time $t$ and $t-1$ as adjacency matrices $A^{(t)}$ and $A^{(t-1)}$, then the differential graph is defined as:

$$\Delta A^{(t)} = A^{(t)} - A^{(t-1)}$$

This captures the change in stock relationships between two consecutive time points. The differential matrix reflects "emerging dependencies" or "diminishing associations" between stocks, which are crucial for financial modeling, as such changes often signal fund flows, sector rotation, or other microstructural shifts in the market.

Building on this idea, we incorporate the differential attention mechanism as an additional structural prior into the computation of self-attention scores:

$$DiffAttention(Q, K, V) = softmax\left(\frac{QK^T}{\sqrt{d_k}} + \lambda \Delta A^{(t)}\right) V$$

Here, $\lambda$ is a learnable weight coefficient that controls the influence of the differential graph on attention allocation. Compared to traditional attention mechanisms that only utilize the current (static) graph structure, differential attention more effectively captures the "relationships in flux" that are most relevant to stock prediction tasks, offering a clear advantage.

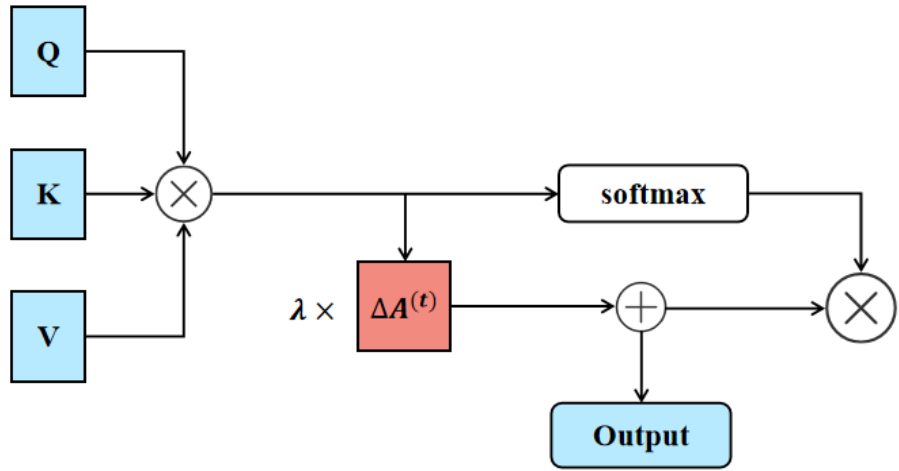

$$DiffAttention(Q, K, V) = softmax\left(\frac{QK^T}{\sqrt{d_k}} + \lambda \Delta A^{(t)}\right)V$$

Figure 4 Simple schematic of differential attention

# 3. Methodology

## 3.1 The basic structure of DGT

The core of DGT lies in its ability to integrate spatial correlations and temporal dependencies, dynamically adjusting the attention mechanism to filter out market noise and uncover key stock relationships. This approach overcomes the over-smoothing and scattered attention weights problems found in traditional Graph Convolutional Networks (GCNs). At the same time, through a flexible graph construction strategy, DGT effectively captures both local and global stock relationships.

Figure 5 compares the basic structures of static and dynamic spatio-temporal graphs:

In the static spatio-temporal graph (Figure 5a), the topology of the graph (i.e., the way nodes are connected) remains fixed across three time points, while only the node attributes change over time.

In contrast, in the dynamic spatio-temporal graph (Figure 5b), both the node attributes and the graph topology evolve over time. Notably, at time step $t_2$, the connections between nodes are visibly different from those at $t_1$.

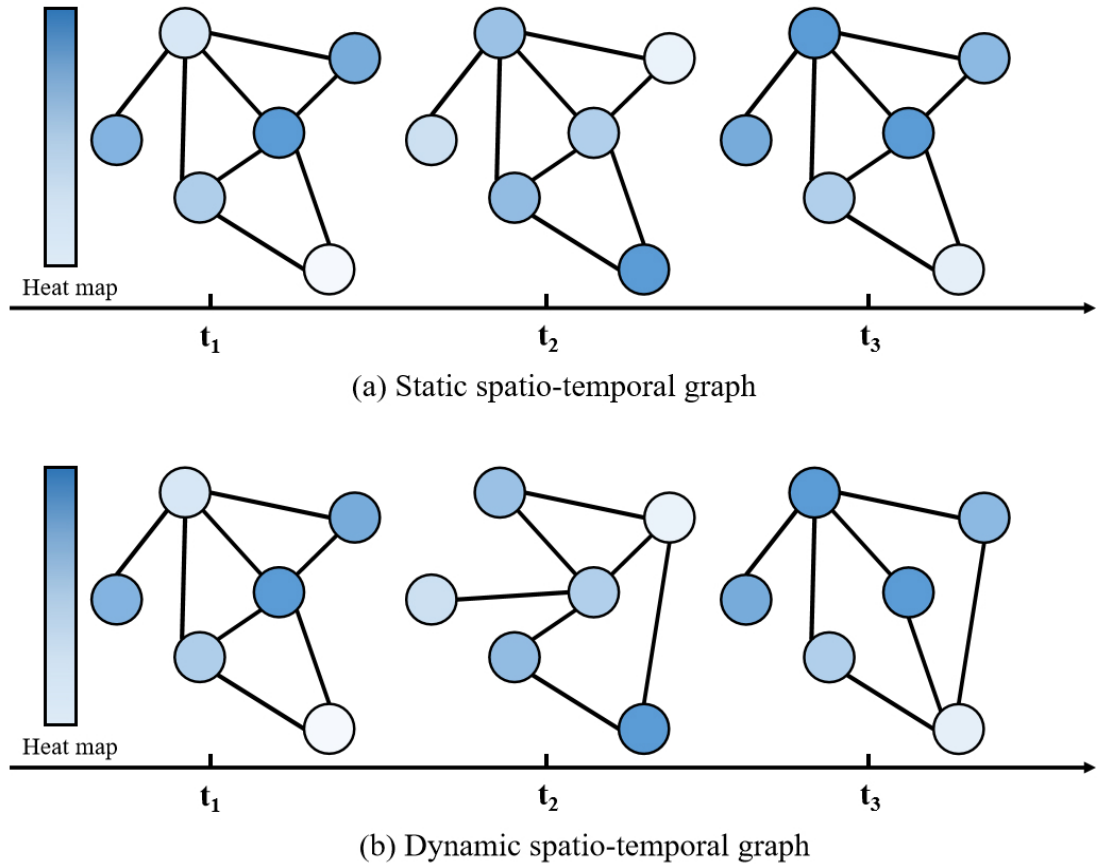

**Figure 5 Basic architecture of static vs. dynamic spatio-temporal graphs**

**Source(s)**: Spatio-Temporal Graph Neural Networks for Predictive Learning in Urban Computing: A Survey (Jin et al., 2019)

Temporal dependencies reflect both the short-term and long-term relationships of stock behavior over time. DGT incorporates the differential attention mechanism introduced in Section 2.3, which dynamically applies spatial correlations at each time step $t$ by generating a dynamic attention matrix. This mechanism uses current market data, combining global and local correlations to adaptively adjust edge weights, thereby alleviating the over-smoothing issue.

In other words, for each time step $t$, DGT follows a two-step workflow: (1) Differential graph attention reevaluates spatial correlations based on market data and generates a dynamic attention matrix. (2) This matrix is then combined with temporal attention to capture causal relationships across time steps.

Through the synergistic interaction between temporal and spatial modeling, the model flexibly combines pre-defined local and global correlation graphs and employs dynamic attention to ensure adaptive correlation thresholds, thus improving the robustness of predictions (Ye et al., 2025).

## 3.2 Input Projection

To effectively fuse stock price data with temporal and stock identity information, this paper introduces an input projection mechanism. Specifically, for each day $t$, we obtain a vector of stock prices:

$$p_t = [p_t^1, p_t^2, \ldots, p_t^N]^T$$

Each component represents the price of stock $s$ on day $t$. Since raw prices are scalar values and insufficient for modeling complex inter-stock relationships and temporal dynamics, we first apply a Feed-Forward Network (FFN) to nonlinearly project the price vector into a d-dimensional embedding space, extracting higher-order features and patterns.

Furthermore, different stocks exhibit different characteristics and long-term stable features. To enable the model to perceive the intrinsic differences among stocks, we introduce a learnable embedding vector for each stock $s$:

$$StockEmbedding(s) \in R^d$$

This helps the model maintain the individuality of each stock and better leverage structural relationships across stocks.

Time is also a critical factor in modeling. Stock prices change over time and may exhibit periodic trends. To capture the sequential and positional information between time steps, we introduce a learnable time embedding:

$$TimeEmbedding(t) \in R^d$$

This provides a unique vector for each time step, helping the model identify temporal dynamics across different periods.

Finally, we sum the price projection, stock embedding, and time embedding to obtain the input embedding for each stock at each time step:

$$X_t^s = FFN(p_t)^s + StockEmbedding(s) + TimeEmbedding(t)$$

## 3.3 Temporal Attention

To model the temporal dependencies of each stock $s$, we use the Transformer structure in the decoder to encode the input embeddings from the past $t$ time steps, aiming to predict the embedding at time step $t+1$.

First, for all time step embeddings $X^s$ of stock $s$, we use learnable matrices $W_{Q^s(h)}$, $W_{K^s(h)}$, $W_{V^s(h)}$ to project them into the query, key, and value matrices $Q^s(h)$, $K^s(h)$, $V^s(h)$, respectively:

$$Q^s(h) = X^s W_{Q^s(h)}, K^s(h) = X^s W_{K^s(h)}, V^s(h) = X^s W_{V^s(h)}$$

A causal mask $M$ is applied such that $M_{i,j} = 0 \: if \: i \geq j$, and $-\infty$ otherwise, to prevent each time step from accessing future information.

Next, for each attention head $h$, we compute attention weights and weighted values using the causal mask, ensuring the model only attends to previous time steps:

$$head^s(h) = Attention\big(Q^s(h)K^s(h), V^s(h)\big) = softmax\left(\frac{Q^s(h)\big(K^s(h)\big)^T}{\sqrt{d/H}} + M\right)V^s(h)$$

The outputs of all heads are concatenated and linearly transformed to obtain the final attention output:

$$X'^s = MultiHead(Q^s, K^s, V^s) = Concat\big(head^s(1), \dots, head^s(H)\big)W_{O^s}$$

Finally, to enhance the model's representation capacity and stabilize training, we add residual connections and layer normalization, followed by a feed-forward network:

$$X'^s = FFN\big(LayerNorm(X'^s + X^s)\big) + X'^s$$

## 3.4 Differential Graph Attention

The Differential Graph Attention mechanism is an innovative extension of the differential attention concept to graph-structured data. By integrating predefined correlation graphs with multi-head self-attention, it enables effective learning over dynamic graph structures. In simple terms, this mechanism builds upon the core idea of differential attention but is optimized for graph data. Traditional attention mechanisms may suffer from noise in graph structures, but differential graph attention leverages prior knowledge in the form of predefined graphs to filter out attention noise and highlight important relationships.

The core idea is to inject the adjacency matrix of a predefined correlation graph into the attention computation as prior knowledge and subtract the dynamically generated attention matrix from it to capture structural changes in the market. This method continues the basic principle of differential attention introduced in Section 2.3, which involves computing the difference between two attention distributions to suppress noise and emphasize key relations. Specifically, the differential attention mechanism calculates the difference between two softmax-based attention matrices, and a key parameter $\lambda$ balances their contributions:

$$\lambda = exp(\lambda_{q1} \cdot \lambda_{k1}) - exp(\lambda_{q2} \cdot \lambda_{k2}) + \lambda_{init}$$

Here, $\lambda_{init}$ is an initial offset that gives the model a starting point. Empirically, $\lambda_{init}$ for the first layer can be set to 0.2, and the model can adjust $\lambda$ during training to fit data-specific characteristics.

In the context of differential graph attention, this mechanism is extended to graph-structured inputs. At each time step $t$, for each attention head $h$, we first compute the query, key, and value representations:

$$[Q_{t1}(h); Q_{t2}(h)] = X_t W_{Q_T(h)}, [K_{t1}(h); K_{t2}(h)] = X_t W_{K_T(h)}, [K_{t1}(h); K_{t2}(h)] = X_t W_{K_T(h)}$$

Here, $X_t$ denotes the input node embeddings at time $t$, and $W_{Q_T(h)}, W_{K_T(h)}, W_{V_T(h)}$ are learnable parameter matrices. Then, the output of each differential graph attention head is calculated as:

$$head_t(h) = \left( softmax\left(\frac{Q_{t1}(h)K_{t1}(h)^T}{\sqrt{d/H}}\right) \odot A_t(h) - \lambda softmax\left(\frac{Q_{t2}(h)K_{t2}(h)^T}{\sqrt{d/H}}\right) \right) V_t(h)$$

Here, $A_t(h)$ is the adjacency matrix of the predefined correlation graph for head $h$ at time $t$, $\odot$ denotes element-wise multiplication (Hadamard product), $\lambda$ is a learnable scalar weight balancing the two attention maps, $d$ is the model dimension, and $H$ is the number of attention heads.

The outputs from all heads are concatenated and linearly projected to obtain:

$$X'_t = MultiHead(Q_t, K_t, V_t) = Concat(head_t(1), head_t(2), ..., head_t(H))W_{O_T}$$

Finally, residual connections and a feed-forward network are applied. $RMSNorm$ denotes Root Mean Square Layer Normalization, and $FFN$ denotes a Feed-Forward Network:

$$X'_t = FFN(RMSNorm(X'_t + X_t)) + X'_t$$

A key advantage of the differential graph attention mechanism is its ability to capture both local and global dependencies simultaneously. Studies have shown that stock correlations in financial markets exhibit complementary patterns at different scales (Ma et al., 2024). By designing this mechanism, we allow different attention heads to focus on different levels of dependency. For example, some heads may use local correlation graphs reflecting intra-industry relationships or short-term interactions, while others may rely on global correlation graphs capturing cross-industry relations or long-term market trends. This design enables the model to dynamically integrate information from multiple scales and enhances its understanding of complex market structures.

## 3.5 Learning Rate Selection and Grid Search Optimization

In the training process of deep learning models, hyperparameter selection plays a critical role in determining model performance. Among them, the learning rate—a key hyperparameter that controls the update step size of parameters—directly affects both the convergence speed and quality of the model. In financial time series forecasting tasks, due to the high noise, non-stationarity, and complex interdependencies of market data, models are particularly sensitive to the choice of learning rate. Our experiments show that both Graph Neural Networks (GNNs) and Differential Graph Attention Models exhibit high sensitivity to learning rate selection: an excessively large learning rate may cause gradient explosion or prevent the model from converging to a good local optimum, while an overly small learning rate may lead to slow convergence, low training efficiency, or getting stuck in suboptimal local minima.

Based on these observations, this study adopts a grid search method to systematically optimize the learning rate. Specifically, based on prior experimental experience, we limit the learning rate search range to {0.01, 0.1}. For each model configuration used in our experiments, we try each value in this set and select the optimal learning rate based on performance metrics on the validation set (using RMSE and MAE in this study). Although simple, this grid search strategy has proven effective in our task, successfully identifying the best learning rate setting for each model configuration.

In terms of code implementation, this process is realized through a nested loop structure: the outer loop iterates over different model configuration combinations (e.g., GNN type, use of spatial information, correlation computation method, etc.), while the inner loop iterates over the predefined learning rate values (0.01 and 0.1). For each configuration and learning rate combination, we first load the corresponding dataset, then train the model for a fixed number of epochs, and evaluate its performance on the validation set.

## 3.6 Loss Function and Evaluation Metrics

In this model, we design a well-crafted loss function to optimize model performance, and use appropriate evaluation metrics to assess prediction accuracy. During training, the final embedding at time step $T$ is projected to predict the price for the next day, $T+1$. The training process adopts teacher forcing, enabling the model to predict the next-day price at each time step in parallel, similar to how a standard decoder operates. The prediction process can be represented through a sequence of mathematical operations:

$$A = StatisticalCorrelations(p)$$

$$X' = TemporalAttention(X) + X$$

$$X'' = DifferentialGraphAttention(X', A) + X'$$

$$\widehat{p_{T+1}} = FFN(X''_T)$$

We use Mean Squared Error (L2 distance) as the loss function to measure the difference between predicted and actual prices:

$$\mathcal{L}(p_{T+1}, \widehat{p_{T+1}}) = \sum_{s=1}^{N} \left(p_{T+1}^s - \widehat{p_{T+1}^s}\right)^2$$

This loss function effectively penalizes large prediction errors, encouraging the model to make accurate forecasts of the price trends for different securities. During optimization, model parameters are adjusted to minimize this loss, thereby improving prediction accuracy.

To comprehensively evaluate model performance, we adopt regression metrics commonly used in financial time series forecasting (Lu et al., 2021), including Root Mean Square Error (RMSE) and Mean Absolute Error (MAE):

$$\mathcal{L}_{RMSE} = \sqrt{\frac{\sum_{s=1}^{N}\left(p_{T+1}^s - \widehat{p_{T+1}^s}\right)^2}{N}}$$

$$\mathcal{L}_{MAE} = \frac{\sum_{s=1}^{N}\left|p_{T+1}^s - \widehat{p_{T+1}^s}\right|}{N}$$

Where $N$ is the number of securities being predicted, and $p_{T+1}^s$ and $\widehat{p_{T+1}^s}$ denote the actual and predicted prices of the $s$-th security at time $T+1$, respectively. RMSE imposes a higher penalty on large errors, making it suitable for evaluating model sensitivity to extreme fluctuations; MAE provides an intuitive measure of prediction error without amplifying the impact of outliers. By using these two complementary metrics, we are able to comprehensively assess the model's prediction performance under different scenarios, providing a solid foundation for future model improvements and applications.

## 4. Research process and results

### 4.1 Dataset Selection and Preprocessing

In this study, we selected the daily closing prices of the S&P 500 component stocks as our dataset, covering a 10-year period from March 2, 2015, to February 28, 2025. Due to the delisting of Catalent (CTLT) and Marathon Oil (MRO) at the end of 2024, their historical data is incomplete. Therefore, these two stocks were removed during the data cleaning stage, resulting in a final selection of 470 component stocks for subsequent analysis. All raw data was sourced directly from Yahoo Finance to ensure consistency and timeliness.

To simulate real-world financial quarters, we divided the entire sample period into 39 time blocks, each approximately 64 trading days long. The first 80% of the data was used for training, the next 10% for validation, and the remaining 10% for testing. Specifically, the first 31 blocks (around 8 years) constituted the training set, the following 4 blocks (about 1 year) were used for validation, and the last 4 blocks (about 1 year) formed the test set. This time-based partitioning ensures no data overlap between training, validation, and testing phases, and enables a reliable assessment of the model's performance on unseen future data.

Following the method proposed by Tian et al. (2023), we applied z-score normalization to the raw closing prices of all stocks:

$$\widetilde{P_{i,t}} = \frac{P_{i,t} - \mu_i}{\sigma_i}$$

where $P_{i,t}$ denotes the closing price of stock $i$ on day $t$, and $\mu_i$ and $\sigma_i$ represent the mean and standard deviation of stock $i$ during the training period, respectively. This step eliminates scale differences across stocks and allows the model to focus on relative price changes, providing standardized inputs for the subsequent forecasting tasks.

## 4.2 Experimental Settings

This study constructed a stock price forecasting framework based on graph neural networks (GNNs) to capture complex inter-stock relationships. To comprehensively evaluate the impact of different correlation metrics on prediction performance, we implemented four classical methods for computing correlations: Pearson correlation, mutual information (MI), Spearman correlation, and Kendall's Tau. These methods characterize inter-stock dependencies from various perspectives, providing multi-dimensional edge weight information for constructing the stock relation graph.

For the temporal dimension of the graph structure, we considered three different scopes of

correlation calculation:

(1) Global correlation, which computes a single correlation matrix using the entire training period;

(2) Local correlation, which calculates a separate correlation matrix for each 64-day "quarter" as defined earlier;

(3) Dual correlation, which incorporates both global and local correlations to provide richer temporal-structural features.

This multi-level correlation design aims to capture both long-term stable relationships and short-term dynamic patterns in the stock market.

Regarding model architecture, we implemented two core networks: a GRU-based baseline model and the proposed DGT model. The DGT model integrates both temporal and spatial attention mechanisms. The temporal attention uses a causal mask to ensure that only historical information is used in prediction, while the spatial attention leverages the aforementioned correlation matrices as prior knowledge to guide cross-stock information propagation. In the spatial attention component, we adopted a multi-head differential attention mechanism, which effectively distinguishes between the positive and negative contributions of correlations to the prediction task. To verify the importance of graph structures, we also designed an ablation experiment with spatial attention disabled.

During training, we used mean squared error (MSE) as the loss function and the Adam optimizer for parameter updates, with the learning rate tuned within the range $\{0.01, 0.1\}$. To improve the model's generalization ability, a many-to-many prediction mode was adopted during training, where the model predicts the next-day price at each time step. In contrast, a many-to-one prediction mode was used during validation and testing, where the model predicts the next-day price based on the past 64 days of price sequences.

For performance evaluation, we used two widely adopted regression metrics in financial time series forecasting: Root Mean Square Error (RMSE) and Mean Absolute Error (MAE). These metrics respectively measure the magnitude and direction of prediction deviations from ground truth.

All experiments were conducted under the same computational environment. The models were trained for 100 epochs, with performance evaluated on the validation set every 10 epochs. The best-performing model parameters on the validation set were saved for final testing. Finally, RMSE and MAE results across all models were compared to assess overall performance.

## 4.3 Experimental Results

We begin by conducting an exploratory data analysis on the S&P 500 dataset. For each of the four correlation measures (global and local versions of Pearson, MI, Spearman, and Kendall's Tau), we visualize the three stocks most correlated with AAPL. The visualizations are as follows:

### 4.3.1 Global Information

**(1) Global Pearson**

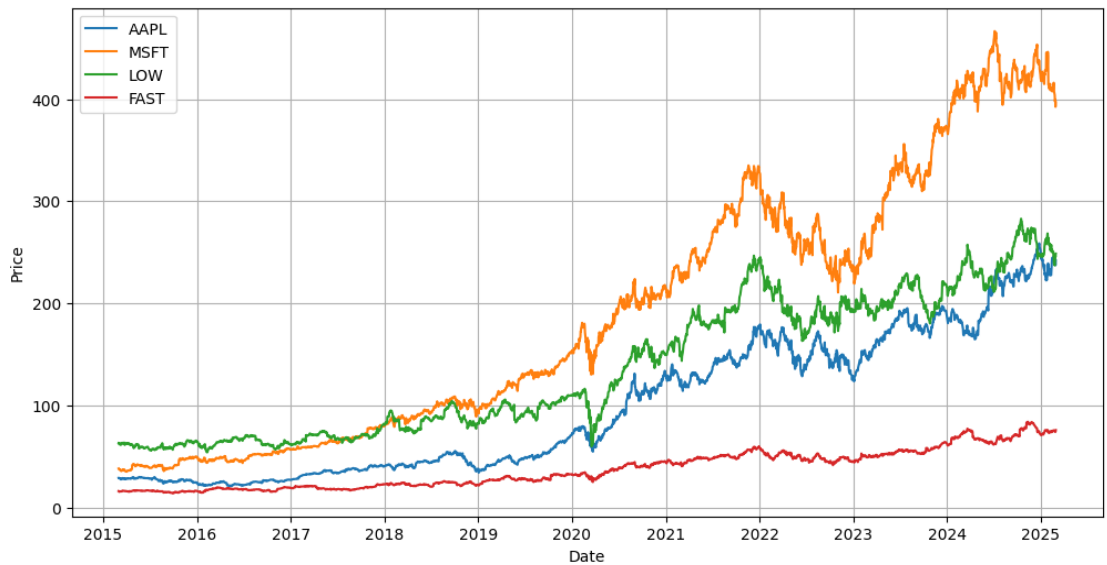

Figure 6 Top 3 stocks most correlated with AAPL under Global Pearson

**(2) Global MI**

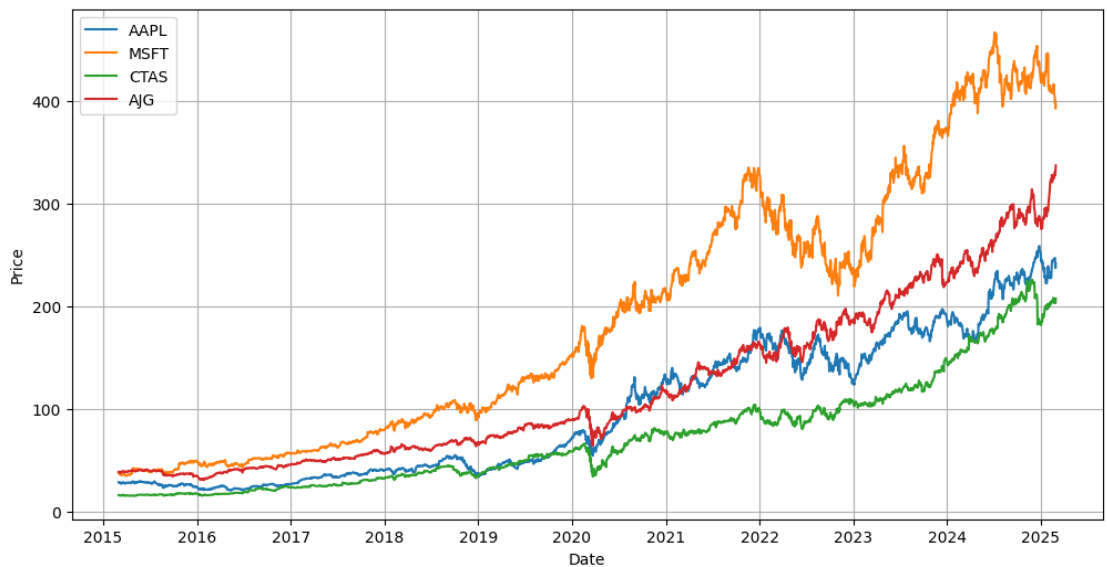

Figure 7 Top 3 stocks most correlated with AAPL under Global Mutual Information

### (3) Global Spearman

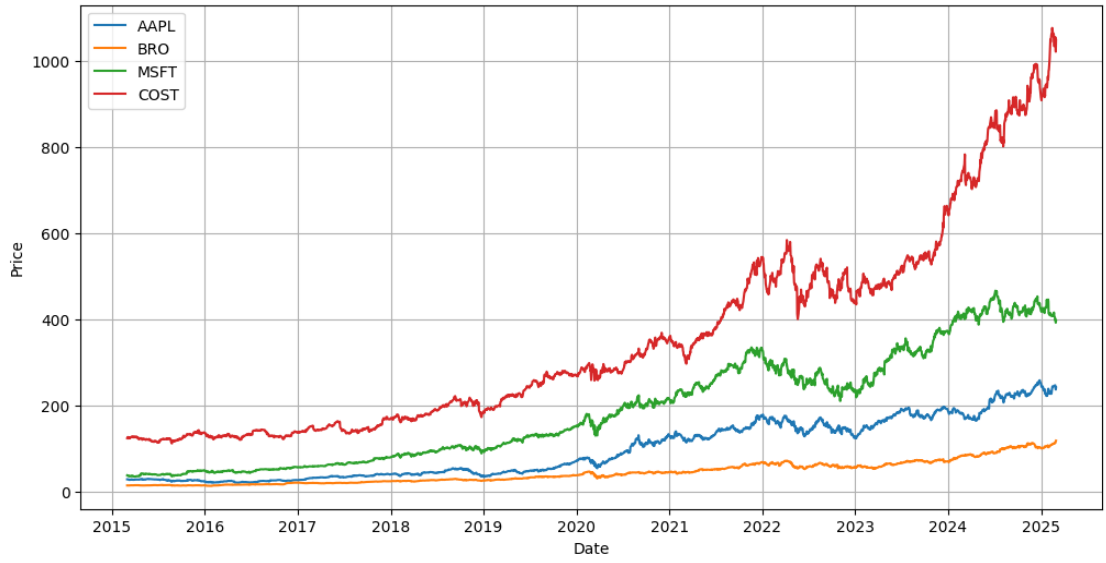

**Figure 8 Top 3 stocks most correlated with AAPL under Global Spearman**

### (4) Global Kendall's Tau

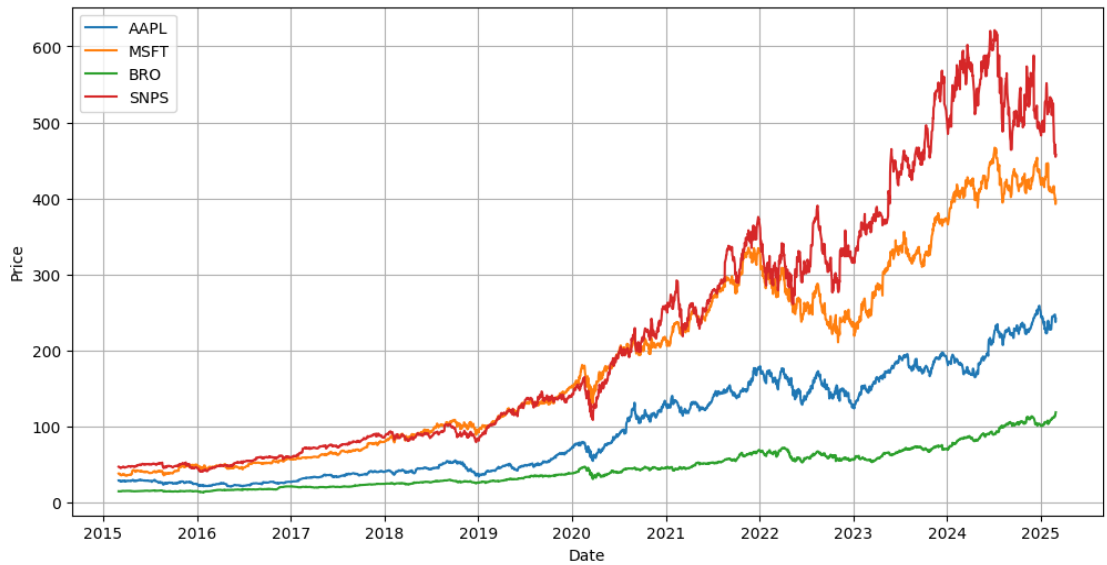

**Figure 9 Top 3 stocks most correlated with AAPL under Global Kendall's Tau**

**Table 1: Top correlated stocks with AAPL under four global correlation methods**

| Correlation method | Top 3 stocks most correlated with AAPL |
|---|---|
| Pearson | MSFT |
|  | LOW |
|  | FAST |
| MI | MSFT |
|  | CTAS |
|  | AJG |
| Spearman | BRO |
|  | MSFT |
|  | COST |
| Kendall's Tau | MSFT |
|  | BRO |
|  | SNPS |

An analysis of AAPL's correlation with other stocks from 2015 to 2025 reveals that Microsoft (MSFT) consistently shows strong correlations across all four methods, indicating similar market performance between the two tech giants. However, the specific set of top three correlated stocks varies slightly depending on the correlation metric used.

Among the methods, global Kendall's Tau and Spearman show superior performance in capturing the true co-movement patterns. As seen in the charts, the stocks identified by these two non-parametric methods exhibit price trends highly consistent with AAPL, especially during major market events such as the COVID-19 crisis in 2020. These methods emphasize directional consistency over magnitude, making them more robust in identifying genuine trend correlations. In contrast, Pearson and MI are more sensitive to absolute price level differences.

A particularly noteworthy case is Costco (COST), whose stock price exceeded $1000 in 2025—much higher than most stocks—yet still displayed a highly similar movement pattern to AAPL, illustrating that correlation measures capture trend patterns rather than price ranges.

### 4.3.2 Local Information

**(1) Local Pearson**

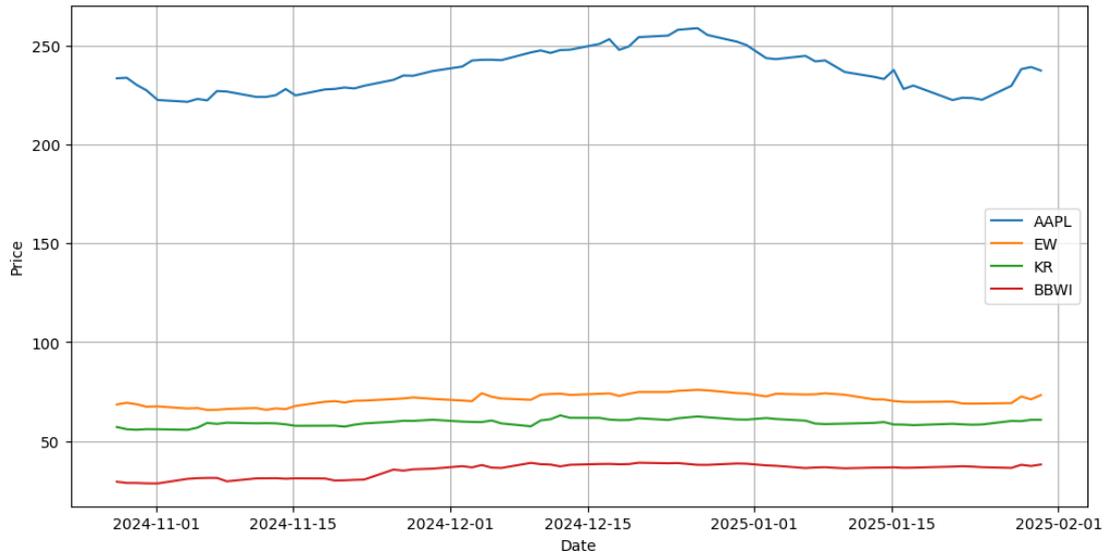

Figure 10 Top 3 stocks most correlated with AAPL under Local Pearson

## (2) Local MI

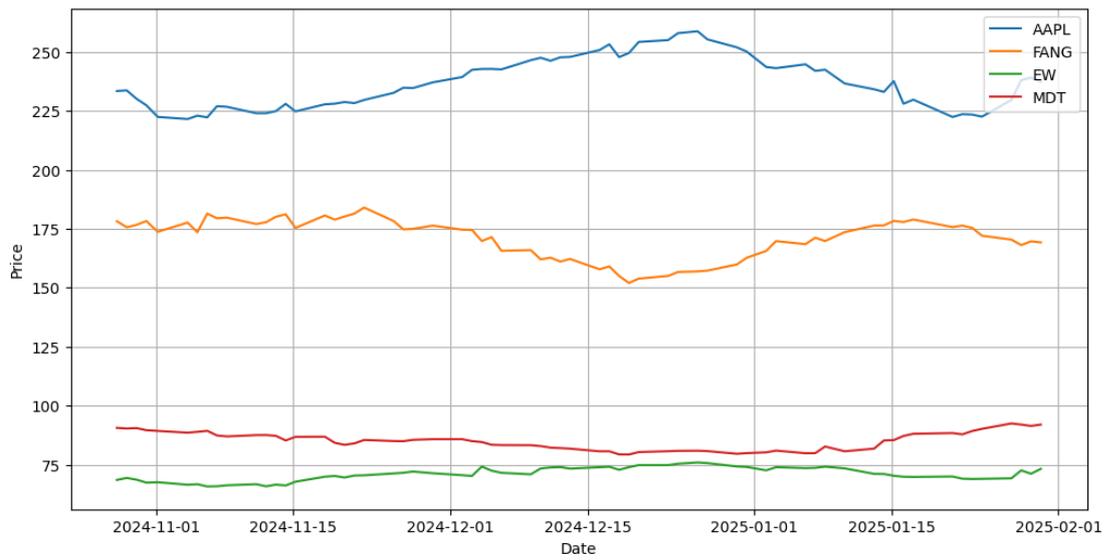

Figure 11 Top 3 stocks most correlated with AAPL under Local Mutual Information

## (3) Local Spearman

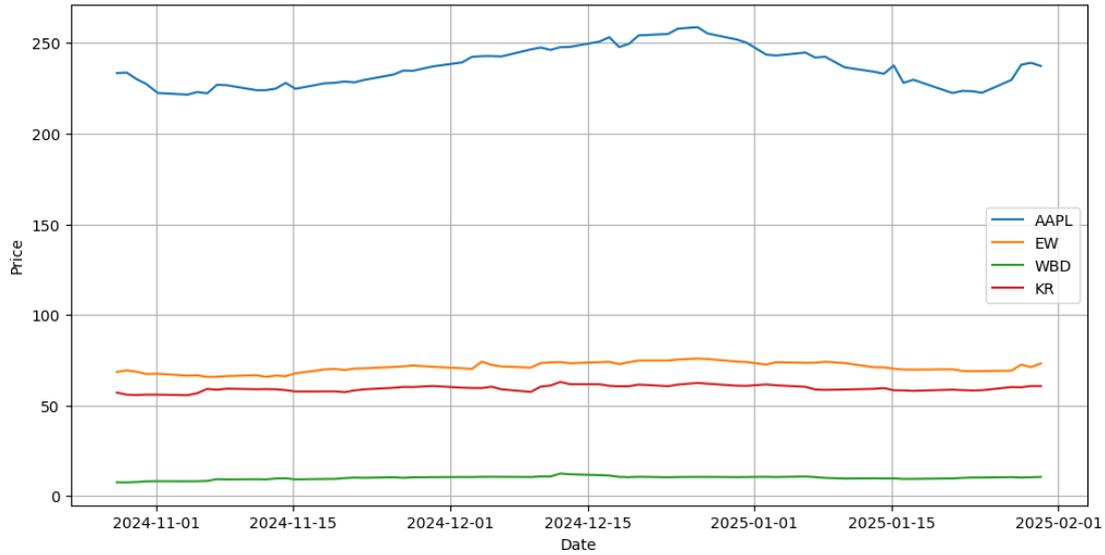

Figure 12 Top 3 stocks most correlated with AAPL under Local Spearman

### (4) Local Kendall's Tau

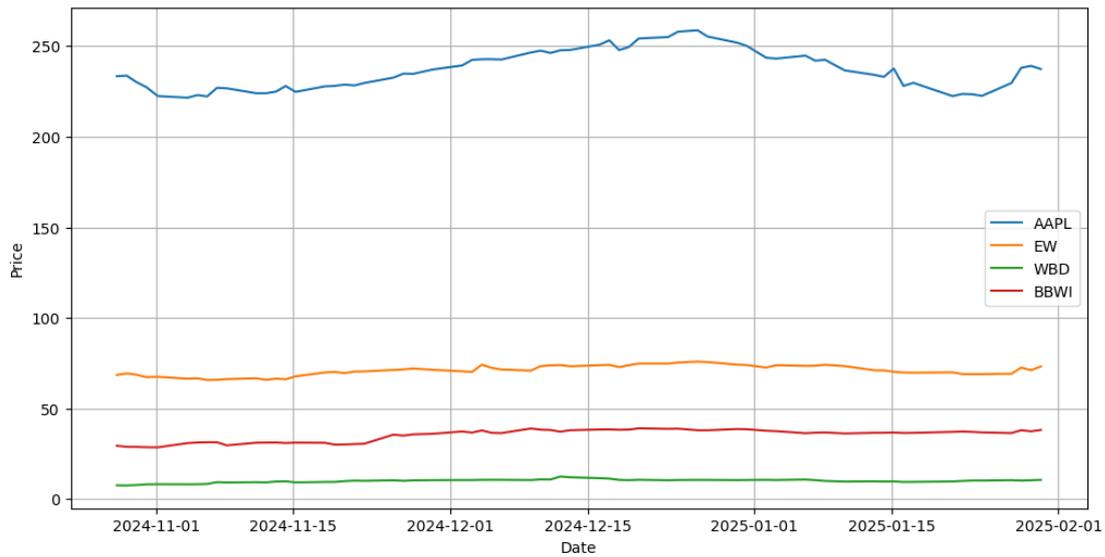

Figure 13 Top 3 stocks most correlated with AAPL under Local Kendall's Tau

Table 2: Top correlated stocks with AAPL under four local correlation methods

| Correlation method | Top 3 stocks most correlated with AAPL |
|---|---|
| Pearson | EW |
|  | KR |
|  | BBWI |
| MI | FANG |
|  | EW |
|  | MDT |
| Spearman | EW |
|              | WBD  |
|              | KR   |
|              | EW   |
| Kendall's Tau | WBD  |
|              | BBWI |

The local correlation analysis reveals that Edwards Lifesciences (EW) consistently ranks as one of the top correlated stocks with AAPL across all four methods. During this period, AAPL's stock underwent a wave of increases followed by corrections, and although EW's price was lower, its short-term fluctuations were strongly synchronized with those of AAPL.

Visual inspection highlights clear differences in the accuracy of each correlation method. Local Kendall's Tau stands out as the best performer, as the stocks it identifies (EW, WBD, and BBWI) closely mirror AAPL's trend—especially during the surge in mid-December 2024 and the decline in January 2025. Spearman ranks second. In contrast, MI performs less effectively: for example, FANG shows an opposite trend to AAPL during the second half of December, suggesting that MI may capture more complex nonlinear dependencies that do not necessarily align with trend similarity.

Unlike global correlations which reflect long-term market relationships, these local correlations offer insights into short-term interactions under specific market conditions—providing valuable guidance for short-term traders. The consistent appearance of EW, a medical device company, as AAPL's top local correlate across all methods challenges traditional industry boundaries and uncovers possible short-term cross-sector market linkages, offering new perspectives for portfolio management and risk control.

### 4.3.3 Model Comparison

**Table 3 Experimental Results of GRU Baseline vs. DGT Variants on the S&P 500 Dataset**

| Architecture | Use Spatial | Correlation | Scope  | RMSE     | MAE      |
|--------------|-------------|-------------|--------|----------|----------|
| GRU          | False       | None        | None   | 2.359901 | 0.477939 |
| DGT          | False       | None        | None   | 1.484602 | 0.336108 |
| DGT          | True        | None        | None   | 0.872668 | 0.305674 |
| DGT          | True        | Pearson     | dual   | 0.739612 | 0.240609 |
| DGT          | True        | Pearson     | global | 0.703439 | 0.213345 |
| DGT | True | Pearson | local | 0.557180 | 0.207669 |
| DGT | True | Kendall's Tau | dual | 0.542374 | 0.172766 |
| DGT | True | MI | local | 0.484439 | 0.176774 |
| DGT | True | Spearman | global | 0.475259 | 0.172522 |
| DGT | True | Spearman | dual | 0.471506 | 0.230972 |
| DGT | True | MI | dual | 0.419014 | 0.184866 |
| DGT | True | Kendall's Tau | local | 0.335979 | 0.113357 |
| DGT | True | Spearman | local | 0.331790 | 0.145324 |
| DGT | True | MI | global | 0.301756 | 0.125020 |
| DGT | True | Kendall's Tau | global | **0.237913** | **0.106069** |

Based on the results shown above:

**(1) The DGT architecture significantly outperforms the GRU baseline model**

As shown in Table 3, even without incorporating spatial information, the DGT model achieves an RMSE of 1.48 and an MAE of 0.34, clearly outperforming the GRU baseline in both metrics. This indicates that the DGT architecture has an inherent advantage in handling stock time series data. Even without additional spatial relational information, it significantly surpasses traditional recurrent neural network architectures.

**(2) Incorporating spatial information greatly improves the performance of the DGT model**

Comparing the DGT model with and without spatial information reveals that introducing spatial information reduces the RMSE from 1.48 to 0.87, an improvement of approximately 41%. This significant enhancement demonstrates the importance of spatial relationships in stock market prediction. It shows that effectively leveraging these relationships helps the model better understand market dynamics and stock price movements.

**(3) Kendall's Tau correlation performs best in the global setting**

Among all correlation measures tested, using global Kendall's Tau yields the best performance, with an RMSE of just 0.24 and an MAE of 0.11. In contrast, although Pearson correlation is commonly used, its global RMSE is 0.70 and MAE is 0.21—significantly worse. This result confirms our earlier observation: Kendall's Tau is better at capturing trend similarities between stocks and is less sensitive to outliers, making it more suitable for financial market data.

**(4) The choice of correlation scope significantly affects model performance, with global generally outperforming local and dual scopes**

When comparing different scopes using the same correlation measure, global correlation typically outperforms both local and dual (combined) correlations. This suggests that long-term global correlation patterns may have greater predictive value in stock market forecasting. The only exception is when using Pearson correlation, where local correlation outperforms global correlation. This may be because linear relationships are more apparent in the short term.

**(5) The optimal model configuration highlights the advantage of integrating deep learning with statistical methods**

The DGT architecture that integrates both spatial-temporal information and global Kendall's Tau correlation achieves approximately 90% lower RMSE and 78% lower MAE compared to the GRU baseline. This dramatic improvement demonstrates that relying solely on the time series of individual stocks is far from sufficient in stock prediction tasks. Effectively integrating spatial relationships in the market and selecting appropriate correlation measures is critical to model performance.

# 5. Discussion

## 5.1 Clustering Method Selection

In stock market analysis, clustering stocks into different categories (e.g., technology, consumer) can significantly improve model interpretability and forecasting performance. The previous experiments discussed different types of stocks together, which limits the ability to explain the prediction differences among categories and makes it difficult to determine which types of stocks the model performs better on. Therefore, this paper introduces an improvement by applying clustering methods to group stocks more reasonably, aiming to reveal differences in prediction results across categories and thus optimize model design.

However, the choice of clustering method directly affects the stability and applicability of the results, so it's important to evaluate the characteristics of different approaches. Initially, we considered three methods: industry classification, unsupervised clustering, and Dynamic Time Warping (DTW). After analysis, we ultimately selected K-means clustering as our method.

Industry classification (based on standards like Wind or CSMAR) groups stocks according to their sector, offering stable and intuitive results. However, this method relies on manual labeling and cannot capture the dynamic features of stock price fluctuations, which may limit the model's ability to explore the data's inherent structure. DTW, on the other hand, measures the similarity of price curve shapes and is good at capturing nonlinear volatility patterns in time series. But its high computational complexity makes it inefficient for large datasets, and thus unsuitable for fast experiments. In contrast, unsupervised clustering groups stocks based on price or volatility features, automatically discovering natural groupings in the data without manual intervention, with low computational cost and reduced human workload.

Among unsupervised methods, K-means clustering stands out due to its simplicity and high computational efficiency. It is especially suitable for the large-scale stock dataset used in this study. By extracting price or volatility features, K-means effectively groups stocks into distinct categories, laying a solid foundation for subsequent category-based predictive analysis.

To sum up, our improved approach is: Based on the clustering results from K-means, we compute the relevant metrics for each category and construct graph structures accordingly. Then, under a unified DGT model architecture, we compare prediction errors across stock categories to evaluate how different forecasting methods perform on each group.

Before applying K-means, we needed to determine the number of clusters. Therefore, we used the silhouette coefficient method and the elbow method to jointly decide.

From the silhouette plot (Figure 14), it's clear that the silhouette score reaches its maximum at $k = 2$, significantly higher than all other values. A higher silhouette score indicates greater intra-cluster similarity and stronger inter-cluster difference, suggesting a clearer and more reasonable clustering structure. In contrast, at $k = 3$, the silhouette score sharply drops to -0.03, indicating an unreasonable clustering structure. While $k = 4$ shows a local peak, the value (0.05) is still far below the score at $k = 2$.

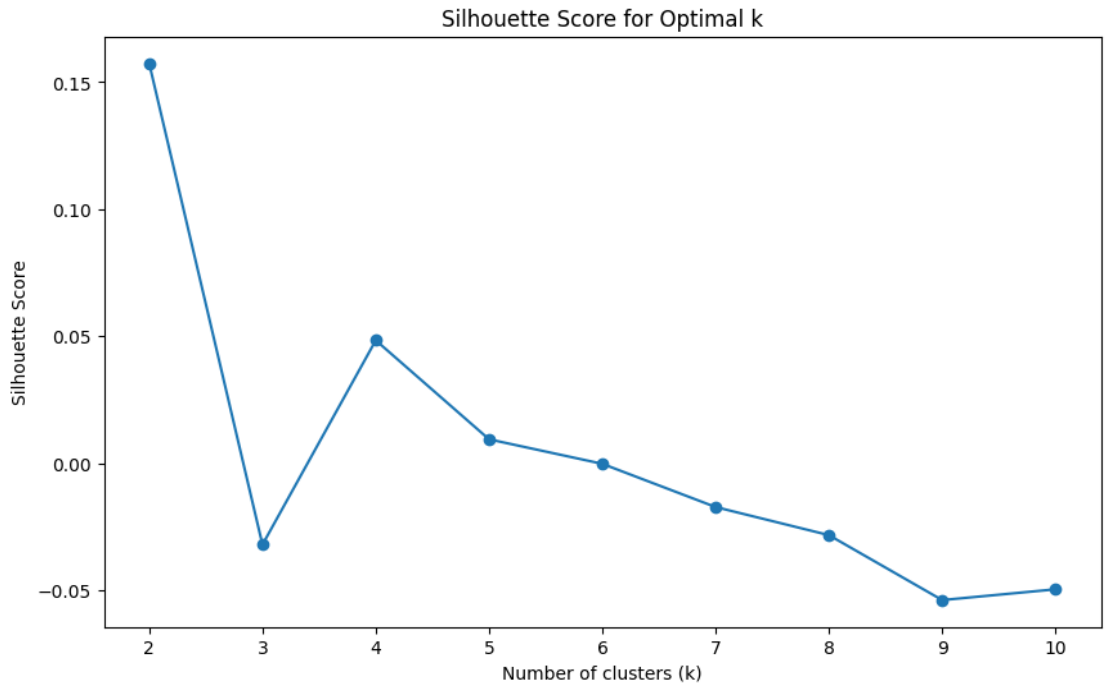

**Figure 14 Silhouette scores for different numbers of clusters**

Although in the elbow plot (Figure 15), k = 2 is not at a typical "elbow" position, the inertia value increases from k = 2 to k = 3, which is unusual and further supports the judgment that k = 3 is unsuitable. A noticeable elbow appears at k = 6, but considering the silhouette results, k = 2 already provides a clear classification, so we ultimately selected k = 2 for the following analysis.

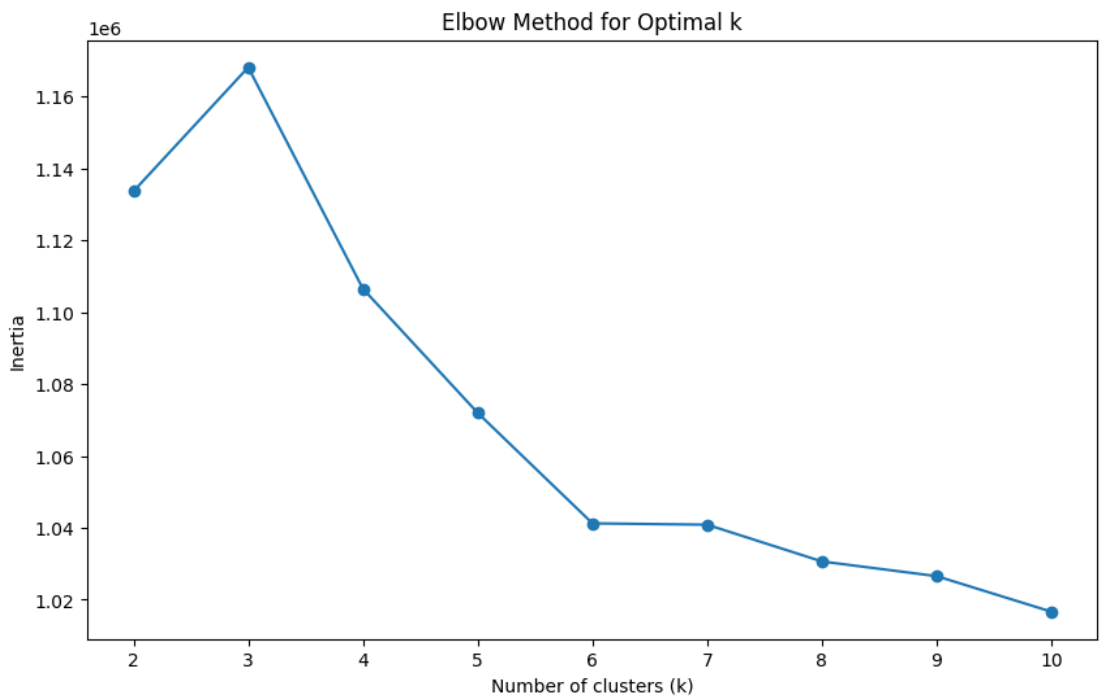

**Figure 15 Inertia values for different cluster numbers (Elbow Method)**

## 5.2 Clustering Results

We applied K-means clustering with k = 2 to the 470 constituent stocks of the S&P 500, resulting in the following cluster scatter plot:

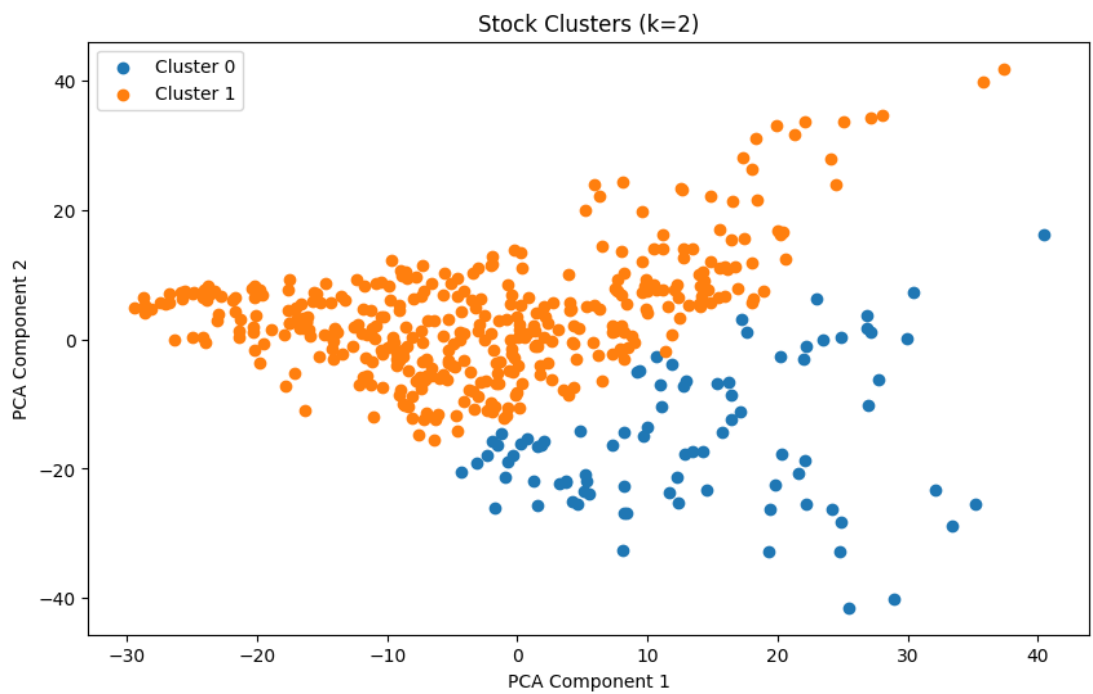

Figure 16 K-means Clustering Result Scatter Plot

The plot shows a good clustering effect with clearly defined boundaries between the two categories. Specifically, K-means divided the stocks into the following two clusters:

**Table 4 Specific Stock Information of cluster 0 and cluster 1**

| Category | Characteristics | Quantity | Specific stock name |
| --- | --- | --- | --- |
| Cluster 0 | High growth/ High volatility | 87 | AAL,AAPL,ADBE,ADI,ADSK,ALB,ALGN,AMAT, AMD,AMZN,ANET,ANSS,APTV,AVGO,AXON,BKNG,BLDR,BX,CCL,CDNS,CRL,CRM,CSGP,CZR,DAL,DECK,DXCM,ENPH,EPAM,EXPE,FCX,FICO,FSLR,FTNT,GNRC,GOOG,GOOGL,IDXX,INTC,INTU,ISRG,JBL,KEYS,KKR,KLAC,LRCX,LULU,LVS,LYV,MCHP,META,MGM,MPWR,MSCI,MSFT,MTCH,MU,NCLH,NFLX,NOW,NTAP,NVDA,NXPI,ON,PANW,PAYC,PODD,PTC,QCOM,RCL,SMCI,SNPS,STX,SWKS,TECH,TEL,TER,TRMB,TSLA,TTWO,TXN,TYL,UAL,URI,WDC,WYNN,ZBRA |
| Cluster 1 | Traditional Blue Chip/ Solid Defense | 383 | A,ABBV,ABT,ACGL,ACN,ADM,ADP,AEE,AEP,AES,AFL,AIG,AIZ,AJG,AKAM,ALL,ALLE,AMCR,AM |

E,AMGN,AMP,AMT,AON,AOS,APA,APD,APH,ARE,ATO,AVB,AVY,AWK,AXP,AZO,BA,BAC,BALL,BAX,BBWI,BBY,BDX,BEN,BF-B,BG,BIIB,BIO,BK,BKR,BLK,BMY,BR,BRK-B,BRO,BSX,BWA,BXP,C,CAG,CAH,CAT,CB,CBOE,CBRE,CCI,CDW,CE,CF,CFG,CHD,CHRW,CHTR,CI,CINF,CL,CLX,CMCSA,CME,CMG,CMI,CMS,CNC,CNP,COF,COO,COP,COR,COST,CPAY,CPB,CPRT,CPT,CSCO,CSX,CTAS,CTRA,CTSH,CVS,CVX,D,DD,DE,DFS,DG,DGX,DHI,DHR,DIS,DLR,DLTR,DOC,DOV,DPZ,DRI,DTE,DUK,DVA,DVN,EA,EBAY,ECL,ED,EFX,EG,EIX,EL,ELV,EMN,EMR,EOG,EQIX,EQR,EQT,ES,ESS,ETN,ETR,EVRG,EW,EXC,EXPD,EXR,F,FANG,FAST,FDS,FDX,FE,FFIV,FI,FIS,FITB,FMC,FRT,GD,GE,GEN,GILD,GIS,GL,GLW,GM,GPC,GPN,GRMN,GS,GWW,HAL,HAS,HBAN,HCA,HD,HES,HIG,HII,HLT,HOLX,HON,HPQ,HRL,HSIC,HST,HSY,HUBB,HUM,IBM,ICE,IEX,IFF,INCY,IP,IPG,IQV,IRM,IT,ITW,IVZ,J,JBHT,JCI,JKHY,JNJ,JNPR,JPM,K,KDP,KEY,KIM,KMB,KMI,KMX,KO,KR,L,LDOS,LEN,LH,LHX,LIN,LKQ,LLY,LMT,LNT,LOW,LUV,LYB,MA,MAA,MAR,MAS,MCD,MCK,MCO,MDLZ,MDT,MET,MHK,MKC,MKTX,MLM,MMC,MMM,MNST,MO,MOH,MOS,MPC,MRK,MS,MSI,MTB,MTD,NDAQ,NDSN,NEE,NEM,NI,NKE,NOC,NRG,NSC,NTRS,NUE,NVR,NWS,NWSA,O,ODFL,OKE,OMC,ORCL,ORLY,OXY,PARA,PAYX,PCAR,PCG,PEG,PEP,PFE,PFG,PG,PGR,PH,PHM,PKG,PLD,PM,PNC,PNR,PNW,POOL,PPG,PPL,PRU,PSA,PSX,PWR,REG,REGN,RF,RJF,RL,RMD,ROK,ROL,ROP,ROST,RSG,RTX,RVTY,SBAC,SBUX,SCHW,SHW,SJM,SLB,SNA,SO,SPG,SPGI,SRE,STE,STLD,STT,STZ,SWK,SYF,SYK,SYY,T,TAP,TDG,TDY,TFC,TFX,TGT,TJX,TMO,TMUS,TPR,TRGP,TROW,TRV,TSCO,TSN,TT,TXT,UDR,UHS,ULTA,UNH,UNP,UPS,USB,V,VLO,VMC,VRSK,VRSN,VRTX,VTR,VTRS,VZ,WAB,WAT,WBA,WBD,WEC,WELL,WFC,WM,WMB,WMT,WRB,WST,WTW,WY,XEL,XOM,XYL,YUM,ZBH,ZTS

**(1) Cluster 0 (87 stocks): High-growth, high-volatility companies**

This cluster focuses on technology innovation and emerging consumption, including prominent stocks such as AAPL, AMD, NVDA, META, MSFT, TSLA, NFLX, LULU, and ENPH. Stocks in this group typically have high intraday return standard deviations and beta values significantly above 1, indicating high sensitivity to market sentiment and sector trends. Their valuation metrics are generally elevated, focusing more on future growth expectations than current cash flows. Under themes like AI, cloud computing, and renewable energy, these stocks often exhibit nonlinear resonance effects.

**(2) Cluster 1 (383 stocks): Traditional blue-chip, stable defensive companies**

This group includes leaders from sectors such as finance (JPM, GS), consumer staples (PG, KO, WMT), healthcare (JNJ, PFE), utilities (DUK, SO), energy (XOM, CVX), industrials, REITs, etc. These stocks have lower return volatility and beta values typically close to or slightly below 1, offering defensive characteristics. Their valuations are reasonable, with stable P/E and P/B ratios, driven by mature businesses and consistent cash flow. They are well-suited for conservative allocation and core defensive asset strategies, showing high synchrony with broader market movements.

**Table 5 Cluster 0 v.s. Cluster 1 Features**

|  | Cluster 0 | Cluster 1 |
| --- | --- | --- |
| Size | 87 | 383 |
| Style | Growth-oriented, high volatility | Value-oriented, defensive |
| Sectors | Tech, AI, Chips, Green energy, Cloud, Media, Discretionary | Finance, Energy, Healthcare, Utilities, Industrials, Real Estate, Staples |
| Investor Type | Theme-driven, long-term growth, risk-tolerant | Conservative, stable returns, institutional |
| Risk | High (sensitive to rates/policy) | Low (stable cash flow) |

## 5.3 Comparison of Prediction Performance between Two Stock Clusters

**Table 6 Comparison of Prediction Errors Between Two Clusters (Sorted by Cluster 0 RMSE in Descending Order)**

| Correlation | Scope | RMSE Cluster 0 | RMSE Cluster 1 | MAE Cluster 0 | MAE Cluster 1 | p-value RMSE | p-value MAE |
| --- | --- | --- | --- | --- | --- | --- | --- |
| Pearson | dual | 1.553260 | 0.351068 | 0.424185 | 0.198909 | 7.14e-39 | 5.61e-110 |
| Pearson | global | 1.509013 | 0.299953 | 0.385628 | 0.174210 | 5.17e-37 | 4.78e-102 |
| Kendall's Tau | dual | 1.153521 | 0.242359 | 0.296565 | 0.144645 | 2.20e-32 | 1.57e-90 |

| Correlation | Scope | RMSE Cluster 0 | RMSE Cluster 1 | MAE Cluster 0 | MAE Cluster 1 | p-value RMSE | p-value MAE |
|---|---|---|---|---|---|---|---|
| Pearson | local | 1.152632 | 0.281393 | 0.344164 | 0.176663 | 4.24e-34 | 4.59e-112 |
| MI | local | 1.026468 | 0.220572 | 0.297737 | 0.149297 | 3.52e-30 | 1.05e-110 |
| Spearman | global | 1.006308 | 0.217139 | 0.278374 | 0.148477 | 7.60e-30 | 4.31e-88 |
| Spearman | dual | 0.926268 | 0.279153 | 0.325019 | 0.209609 | 4.68e-28 | 3.80e-86 |
| MI | dual | 0.849035 | 0.227396 | 0.284817 | 0.162161 | 1.33e-27 | 1.28e-113 |
| Kendall's Tau | local | 0.714265 | 0.150449 | 0.185772 | **0.096908** | 9.23e-20 | 3.03e-81 |
| Spearman | local | 0.681601 | 0.171928 | 0.217030 | 0.129036 | 7.87e-20 | 2.13e-90 |
| MI | global | 0.613281 | 0.162188 | 0.181148 | 0.112271 | 1.86e-17 | 7.44e-68 |
| Kendall's Tau | global | **0.480719** | **0.130257** | **0.142959** | 0.097689 | 6.45e-11 | 1.91e-48 |

The results in the table clearly demonstrate that the DGT model exhibits significant differences in prediction performance across the two stock clusters. Among the correlation measures used, Kendall's Tau delivers the best prediction performance, and models built using global correlation outperform those using local or hybrid (dual-scope) correlations.

**(1) Higher Prediction Error for High-Volatility Stocks**

The model performs noticeably differently across the two clusters. From the overall error metrics, Cluster 1 stocks (typically traditional blue-chip or defensive stocks with stable behavior) are modeled more accurately than Cluster 0 stocks (high-growth and high-volatility stocks). The model appears better suited for predicting conservative and stable stocks.

For Cluster 1, across all combinations of correlation measures and scopes, RMSE consistently falls between 0.13 and 0.35, and MAE between 0.10 and 0.21, indicating that the return series is smoother and its comovement structure is more easily captured by linear models.

In contrast, Cluster 0 stocks show significantly higher prediction errors, with RMSE ranging from 0.48 to 1.55, and MAE between 0.18 and 0.42. This result aligns with the nature of high-growth, high-volatility stocks, whose price movements are more sensitive to market sentiment and thematic factors. These stocks exhibit complex nonlinear comovement structures, posing greater challenges for traditional forecasting models.

**(2) Nonlinear Correlation Measures (Kendall & MI) Deliver Better Performance**

Within Cluster 0, the choice of correlation measure significantly affects model performance. Among the methods compared, Kendall's Tau consistently yields the best results. In particular, under the global correlation setting, it achieves the lowest RMSE of 0.481 and MAE of 0.098, outperforming other methods by a clear margin.

Mutual Information (MI) ranks second, also demonstrating strong predictive power under global correlation settings due to its capacity to capture nonlinear dependencies. Spearman's rank correlation performs moderately well, while Pearson's linear correlation yields the worst results among all methods, especially under the hybrid scope, where RMSE reaches 1.55 and MAE 0.42, indicating a failure to effectively capture the intricate comovements among high-growth stocks.

In Cluster 1, Kendall's Tau also performs well, both in global and local correlation settings.

**(3) Global Correlation Networks Are More Suitable for Modeling High-Growth Stocks**

The choice of scope also has a noticeable impact on prediction performance. In Cluster 0, global correlation consistently outperforms local and hybrid scopes. For instance, using Kendall's Tau under the global setting results in significantly lower RMSE compared to local or hybrid settings. This suggests that in high-volatility stock groups, a broader and more stable global correlation network helps the model better capture systemic comovement signals.

Local or hybrid correlations may reflect short-term interactions, but in high-growth stocks, these tend to be transient and theme-driven, which may hinder stable model fitting.

To sum up, traditional blue-chip stocks, with their strong stability and linear comovements, are more amenable to prediction using simple linear correlation methods. On the other hand, high-growth stocks require more nuanced modeling of nonlinear structures. For the latter, adopting nonlinear correlation measures such as Kendall's Tau or Mutual Information, combined with global network structures, is key to improving prediction accuracy. This conclusion also provides theoretical and empirical support for developing tailored models for different types of assets.

The following figures visualize the DGT model predictions based on different correlation types and scopes for the two representative stocks: AAPL (Apple Inc.) from Cluster 0 and BAC (Bank of America) from Cluster 1, allowing for a more intuitive comparison of how each correlation method affects prediction performance.

**(1) Pearson-Based Predictions**

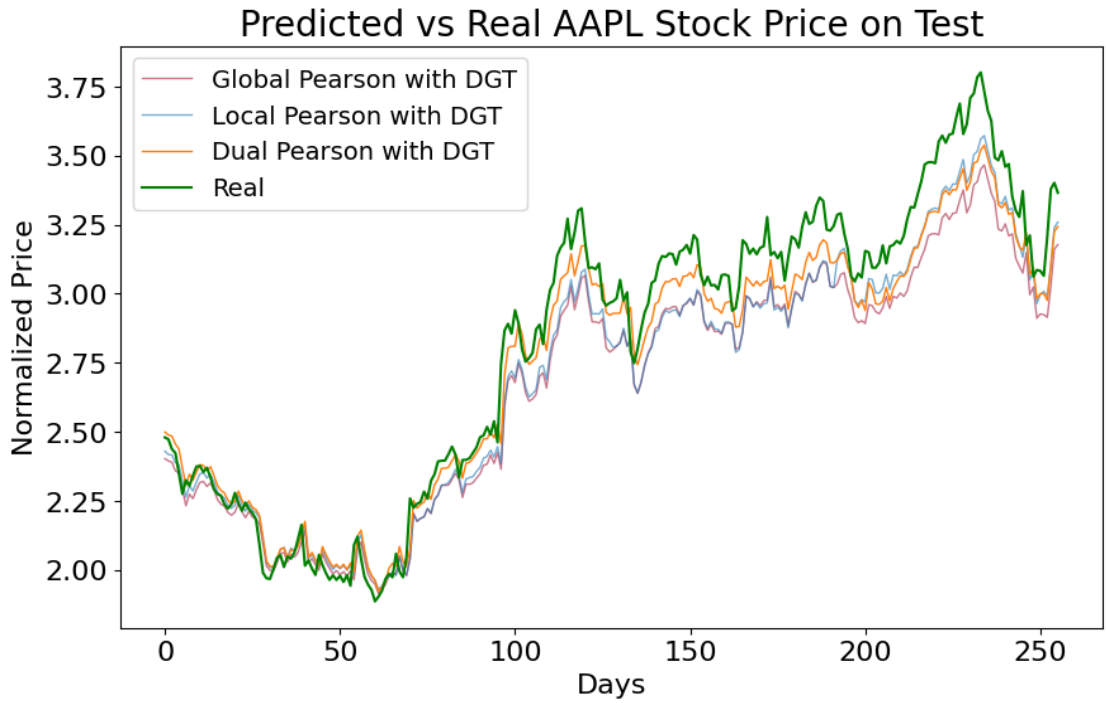

Figure 17 DGT Predictions for Apple Inc. Using Pearson Correlation

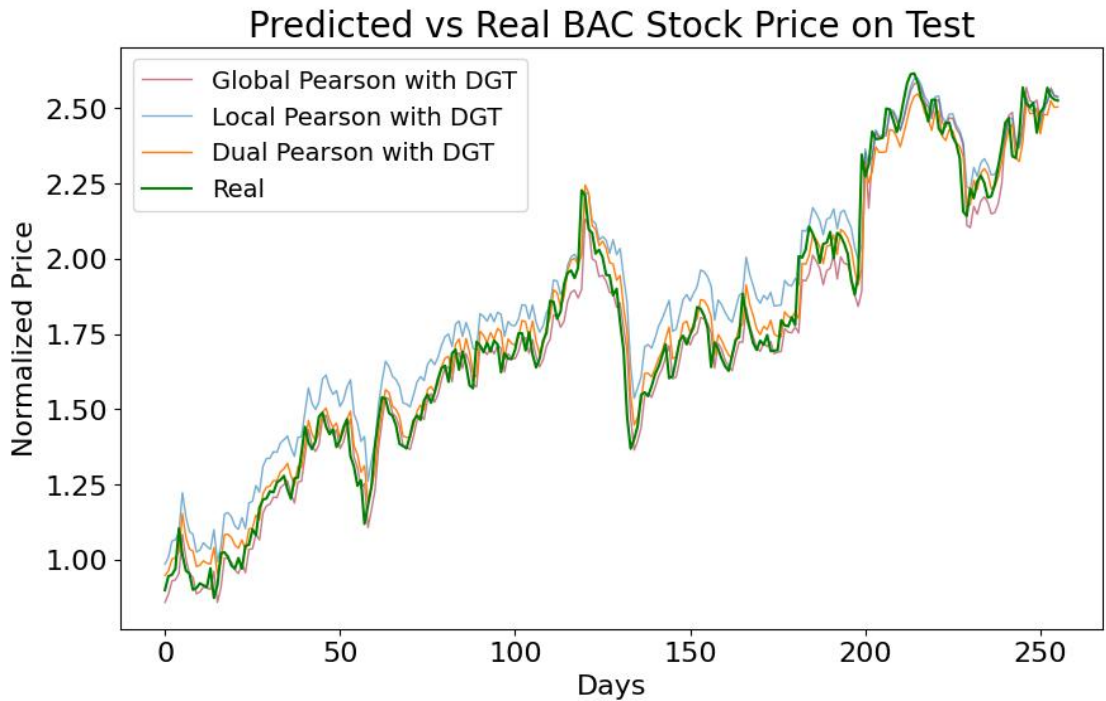

Figure 18 DGT Predictions for Bank of America Using Pearson Correlation

**(2) Mutual Information-Based Predictions**

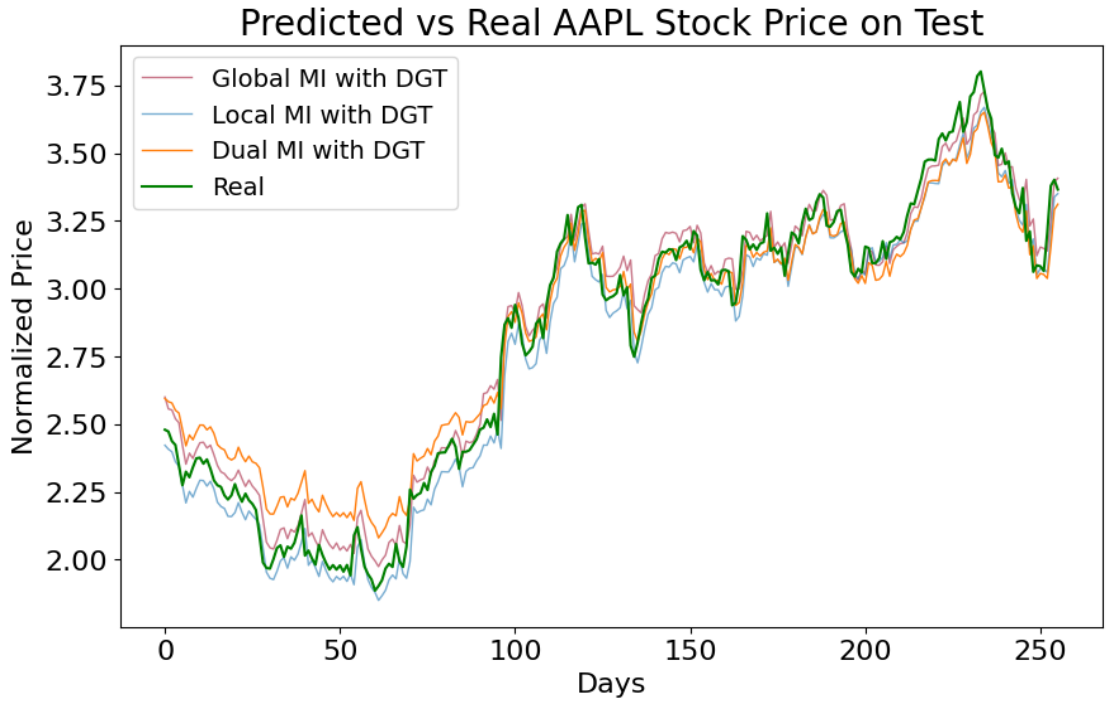

**Figure 19 DGT Predictions for Apple Inc. Using Mutual Information**

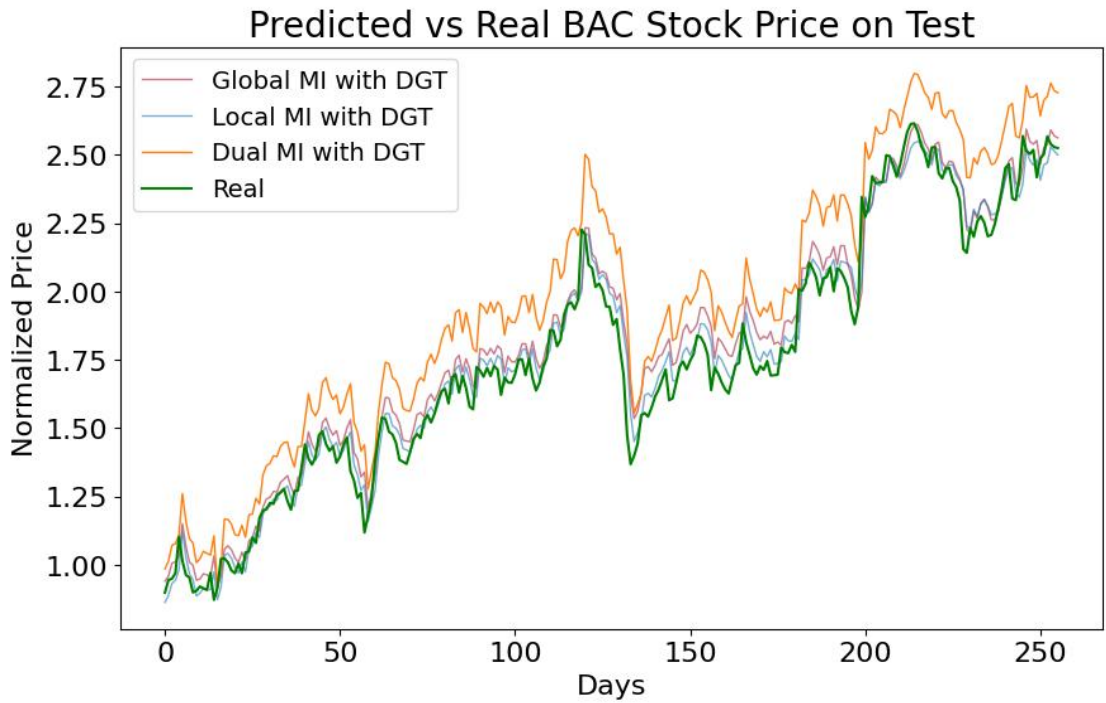

**Figure 20 DGT Predictions for Bank of America Using Mutual Information**

**(3) Spearman-Based Predictions**

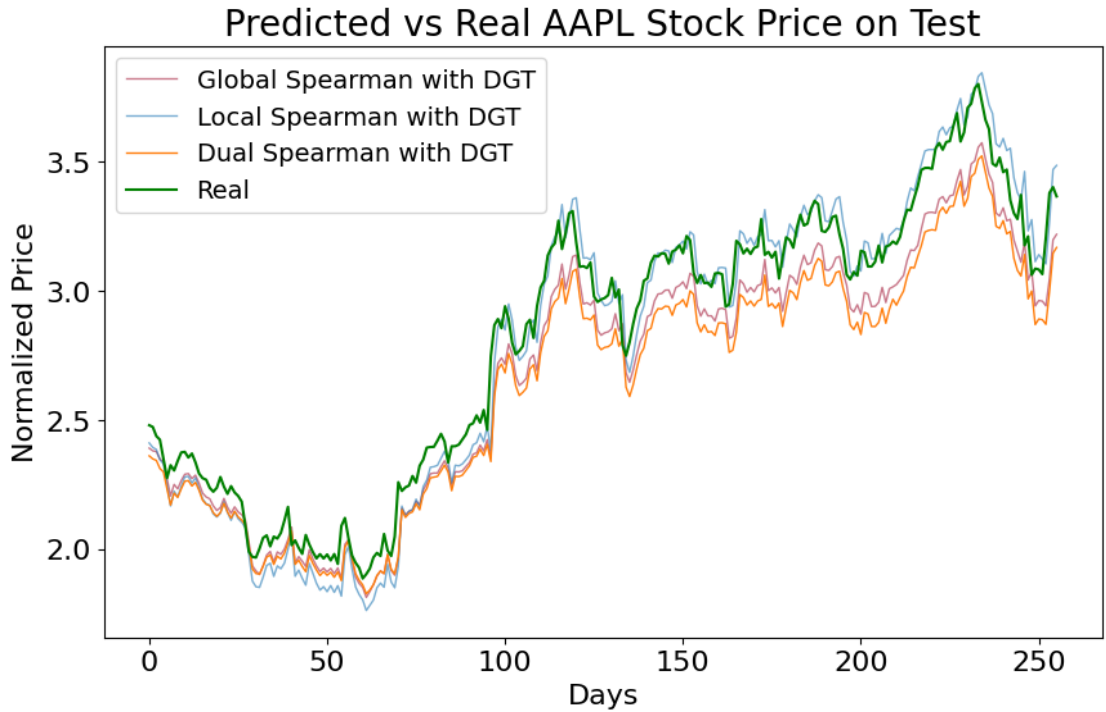

Figure 21 DGT Predictions for Apple Inc. Using Spearman Correlation

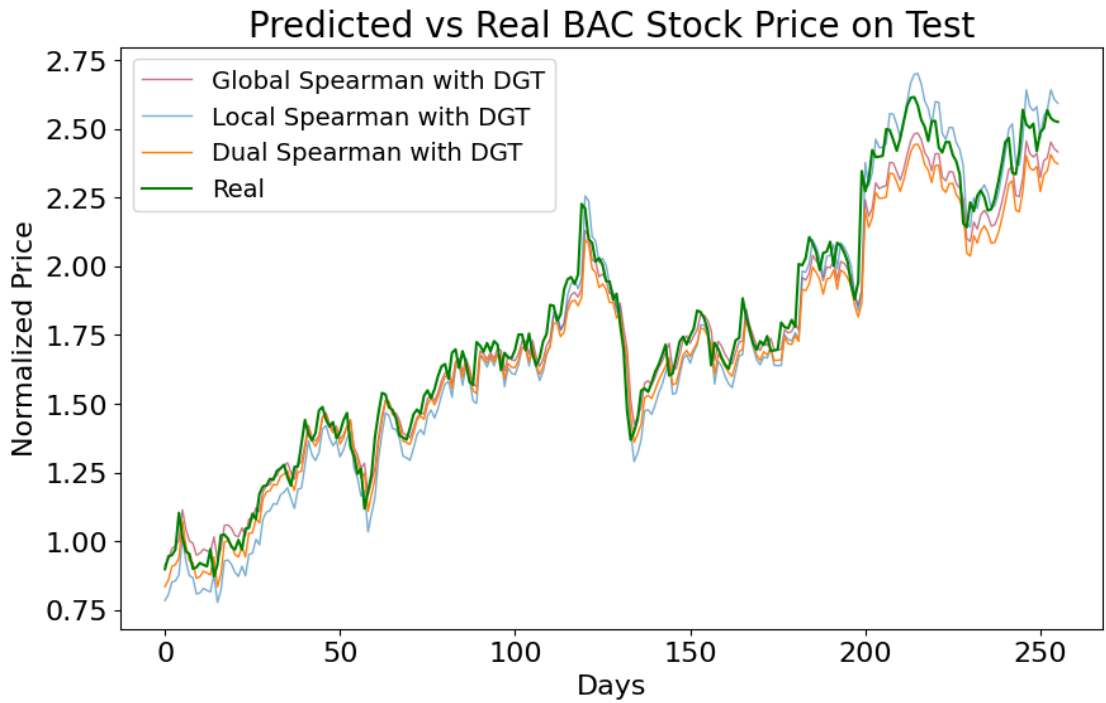

Figure 22 DGT Predictions for Bank of America Using Spearman Correlation

**(4) Kendall's Tau-Based Predictions**

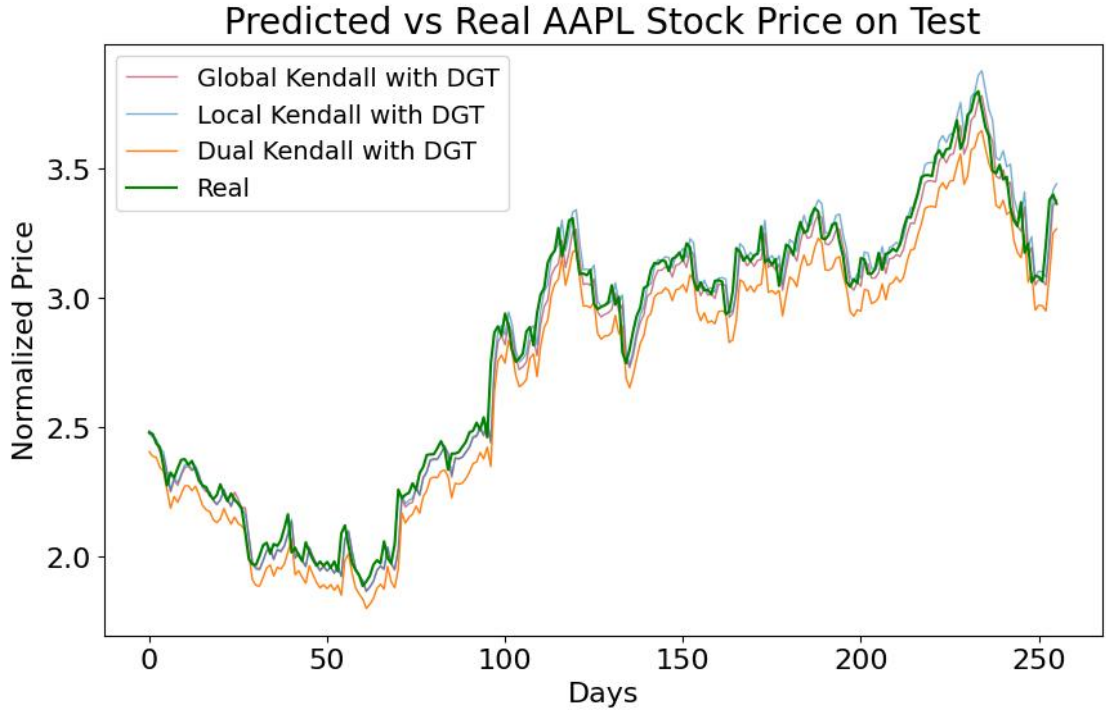

Figure 23 DGT Predictions for Apple Inc. Using Kendall's Tau

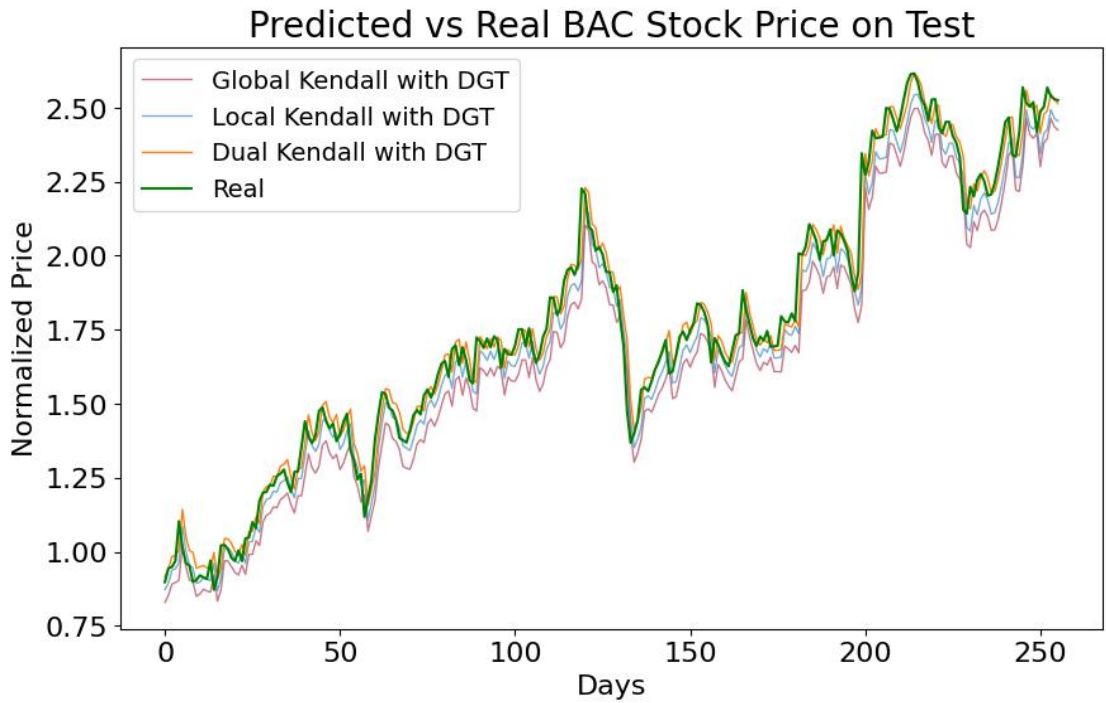

Figure 24 DGT Predictions for Bank of America Using Kendall's Tau

# 6. Conclusion

## 6.1 Main Research Conclusions

This study proposes a dynamic stock relationship modeling framework based on the Differential

Graph Transformer (DGT), significantly improving the prediction accuracy for the prices of S&P 500 constituent stocks. It offers a novel perspective for time-series forecasting and relationship modeling in financial markets. Experimental results demonstrate that, after integrating spatial information, the DGT model shows substantial performance improvements compared to the baseline GRU model. Notably, when using Kendall's Tau as the correlation metric within a global correlation context, the model achieves its best performance—RMSE and MAE drop to 0.24 and 0.11 respectively, representing improvements of approximately 90% and 78% over the baseline. This not only validates DGT's powerful ability to capture dynamic inter-stock relationships and temporal dependencies but also highlights the superiority of nonlinear correlation measures in complex market structures.

In further analysis, the study employs K-means clustering to divide 470 constituent stocks into two groups: high-growth/high-volatility stocks and traditional blue-chip/defensive stocks. The results reveal heterogeneous prediction performances of the model across these clusters. DGT performs especially well on traditional blue-chip stocks, with significantly lower RMSE and MAE compared to high-volatility stocks. This aligns with the lower volatility and stronger linear co-movement characteristics of blue-chip stocks, indicating that high volatility and nonlinear dynamics pose greater challenges to prediction models.

Moreover, the study systematically compares several correlation metrics—Pearson, MI (Mutual Information), Spearman, and Kendall's Tau—across different scopes: global, local, and hybrid. It finds that global correlation networks generally outperform local or hybrid strategies, especially in high-volatility sectors, by more effectively capturing systemic co-movement signals. This suggests that stable long-term relational patterns are crucial for improving prediction accuracy, while dynamic short-term relationships, although informative, may introduce uncertainty in noisy environments. Visual analyses of representative stocks AAPL and BAC further validate the reliability of Kendall's Tau in capturing trend similarity, offering investors a more accurate tool for price forecasting.

## 6.2 Research Limitations

Despite the promising advancements made in predicting S&P 500 prices, this study has several notable limitations:

(1) **Limited Input Features**: The model relies mainly on historical price sequences and does not

incorporate fundamental indicators, macroeconomic variables, or market sentiment data. These additional factors are potentially vital for a more comprehensive understanding of market dynamics, especially during unexpected events or style rotations. Sole dependence on price data may limit the model's ability to capture complex driving forces.

(2) **Market and Time Range Restriction**: The experiments focus solely on the S&P 500 constituents, using data from the U.S. market between 2015 and 2025. The model's generalizability to other markets or different time periods (e.g., bull vs. bear markets) remains untested.

(3) **Simplified Clustering Criteria**: The clustering analysis uses only price-based features to divide stocks, without considering industry classifications, company size, or other latent influencing factors. This may weaken the interpretability and representativeness of the clustering outcomes.

## 6.3 Future Research Directions

Overall, this study advances the frontier of integrating graph neural networks with Transformer architectures and offers empirical support for quantitative investment strategies. The DGT model demonstrates strong practical and theoretical value in financial forecasting. Future research can build upon this work in several directions to overcome existing limitations and further enhance its utility and rigor:

(1) **Incorporating Diverse Data Sources**: Future models could include a wider range of data, such as firm-level financial indicators, news sentiment analysis, or social media signals, to provide a more holistic view of market dynamics.

(2) **Graph Construction Innovations**: Exploring alternative graph construction strategies, such as dynamic thresholding or adaptive graph learning, could improve sparsity and temporal adaptability, reduce dependence on predefined correlation metrics, and enhance responsiveness to structural shifts.

(3) **Cross-Market Validation**: Applying the model to other stock markets (e.g., China A-shares, Hong Kong stocks) or different historical phases could help test its generalizability and robustness, providing empirical evidence for its universal applicability.

(4) **Improving Modeling for High-Growth Stocks**: Given the prediction difficulty of high-growth, high-volatility stocks, future research could focus on developing techniques better suited for nonlinear, volatile data. This includes introducing more sophisticated nonlinear correlation measures or adopting advanced graph neural architectures to improve the model's ability to capture complex co-movement structures.

**Authorship contribution statement:**

Linyue Hu: Writing – original draft, Visualization, Validation, Methodology.

Qi Wang: Writing – review & editing, Supervision, Methodology.